\documentclass[twocolumn,nofootinbib]{revtex4-1}
\usepackage{verbatim}
\usepackage{graphicx}
\usepackage[export]{adjustbox}
\usepackage{float}
\usepackage{amssymb}
\usepackage[utf8]{inputenc}
\usepackage[english]{babel}
\usepackage{amsmath}
\usepackage{mathtools}
\usepackage{amsfonts}
\usepackage{stackrel}
\usepackage{cases}
\usepackage{bm}%
\usepackage{hyperref}
\usepackage{tikz}
\usetikzlibrary{calc}
\usetikzlibrary{decorations.pathreplacing}
\usetikzlibrary{decorations}
\usetikzlibrary{arrows,decorations.markings} 

\newcommand{\be}{\begin{align}}
\newcommand{\ee}{\end{align}}
\newcommand{\dif}{\mathrm{d}}
\newcommand{\R}{\mathbb{R}}
\newcommand{\C}{\mathbb{C}}

 \newcommand{\im}{\mathbf{i}}
 \newcommand{\zmode}{\varphi_0}

 \newcommand{\abs}[1]{\left\vert#1\right\vert}
 \newcommand{\defeq}{\stackrel{\text{def}}=}
 \newcommand{\vertex}{\mathcal{V}}
  \newcommand{\overlap}{\mathsf{q}}
 \usepackage{color}
\usepackage{verbatim}

\newcommand{\red}[1]{{\color{red} #1}}

\definecolor{blue}{rgb}{0,0,1}
\definecolor{green}{rgb}{0,.75,0}
\definecolor{red}{rgb}{0,0,0}
\definecolor{vio}{rgb}{1,0,1}
\definecolor{uv}{rgb}{0.5,0,0.5}
\definecolor{ama}{rgb}{0.3,0.3,0.3}

\begin{document}
\title{Liouville field theory and log-correlated Random Energy Models}
 \author{Xiangyu Cao}\affiliation{LPTMS, CNRS, Univ. Paris-Sud, Université Paris-Saclay, 91405 Orsay, France}\author{Pierre Le Doussal}\affiliation{CNRS-Laboratoire de Physique Théorique de l'\'Ecole Normale Supérieure, 24 rue Lhomond, 75231 Paris Cedex, France}\author{Alberto Rosso}\author{Raoul Santachiara}\affiliation{LPTMS, CNRS, Univ. Paris-Sud, Université Paris-Saclay, 91405 Orsay, France}
 \begin{abstract}
 An exact mapping is established between the $c\geq25$ Liouville field theory (LFT) and the Gibbs measure statistics of a thermal particle in a 2D Gaussian Free Field plus a logarithmic confining potential. 
 The probability distribution of the position of the minimum of the energy landscape is obtained exactly by combining the conformal bootstrap and one-step replica symmetry breaking methods. Operator product expansions in LFT allow to unveil novel universal behaviours of the log-correlated Random Energy class. High precision numerical tests are given. 
 \end{abstract}
 \maketitle 

The problem of a thermal particle embedded in a log-correlated random potential (log-REMs) plays a key role in many physical systems ranging from 2D localisation \cite{kogan96prelocalised,chamon1996localization,castillo97dirac,horovitz2002freezing} to spin-glasses \cite{carpentier2001glass,fyosom2007,fyodorov2007explicit,*fyodorov2008multiscale}, branching process  \cite{derrida1988polymers,krapivsky00kpp,brunet2011branching,arguin2013extremal,aidekon2013branching}, and random matrices \cite{fyo12zeta,*FyoKeat14,fyodorov16char,ABB2015,arguin2015maxima,PaqZeit2016,ChMadNaj2016}.
As a result of the competition between the deep minima of the log-potential and the entropic spreading of the particle, the system undergoes a second order {\it freezing transition} between a high-temperature delocalized phase and a low-temperature glassy phase where the particle is frozen in few minima \cite{carpentier2001glass,derrida1988polymers}.  In the simplest realization of such disordered systems, the random potential is sampled from a 2D Gaussian free field (2D GFF). This allowed for exact predictions of free energy and Gibbs measure statistics \cite{fyodorov08rem}, in cases where the particle is restricted to simple 1D curves drawn on the 2D GFF potential \cite{fyodorov2009statistical,fyodorov2010freezing,fyodorov2016interval,cao15gff,cao16maxmin,ostrovsky2009mellin,ostrovsky2016gff}.  Unfortunately, no results are known in 2D, despite powerful tools of integrability and conformal field theory, \textit{e.g.} the Dotsenko-Fateev integrals \cite{dotsenko1984conformal} generalizing the Selberg integrals used for 1D curves. 

One of the most studied 2D conformal field theories is the Liouville field theory (LFT) that describes the 2D quantum gravity \cite{polyakov1981quantum,zamolodchikov2007lectures,teschner01LFT}, \red{and plays an important rôle in the holography correspondence with $(2+1)$-D gravity, see \textit{e.g.} \cite{carlip05gravity,Alkalaev2016holographic} and references therein}. Although LFT is an \textit{interacting} theory, it has strong connections to the 2D GFF. Indeed, this is a general feature of conformal field theories as it is manifest, for instance, in the Coulomb gas approach to critical statistical models \cite{di1987relations,dotsenko1984conformal}. This viewpoint underlies also recent mathematical developments \cite{duplaniter09dual,duplantier2014critical,david2014liouville,kupiainen2015conformal}. 

\begin{figure}
\includegraphics[scale=.6]{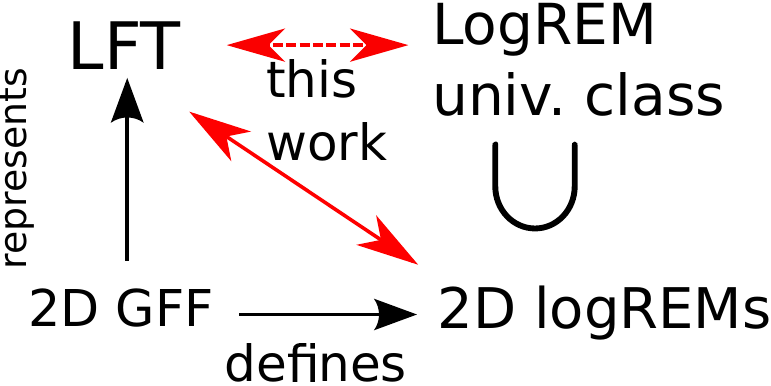}
\caption{The map of connections considered in this work. Building on known relations between LFT and 2D GFF, we establish exact mappings between LFT and logREMs defined by 2D GFF (eq. \eqref{eq:mainhighT} and \eqref{eq:general}). Then we exploit the universality of the logREM class to extend Liouville OPE predictions to all logREMs, \textit{e.g.} eq. \eqref{eq:overlap}.}\label{fig:outline}
\end{figure}

\begin{figure}[h]
\includegraphics[scale=.4,valign=t]{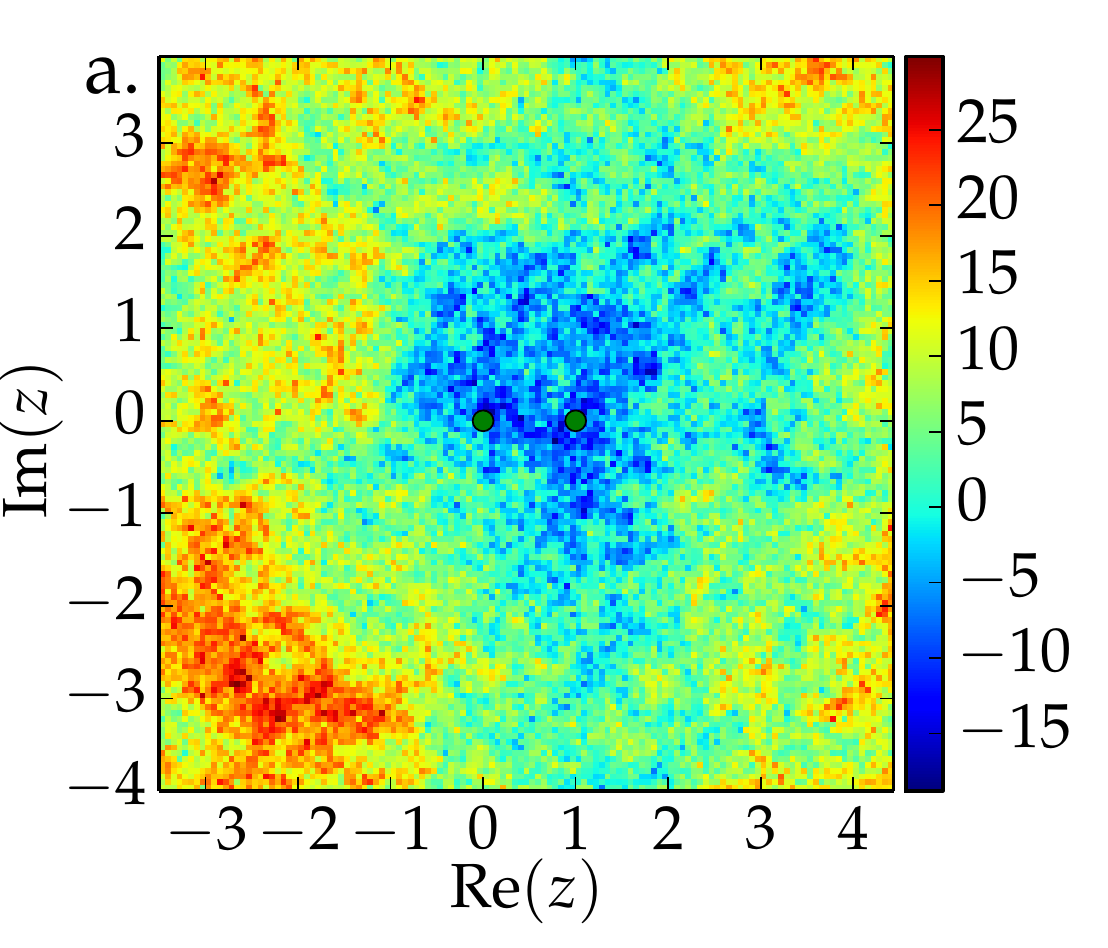} 
\includegraphics[scale= .5,valign=t]{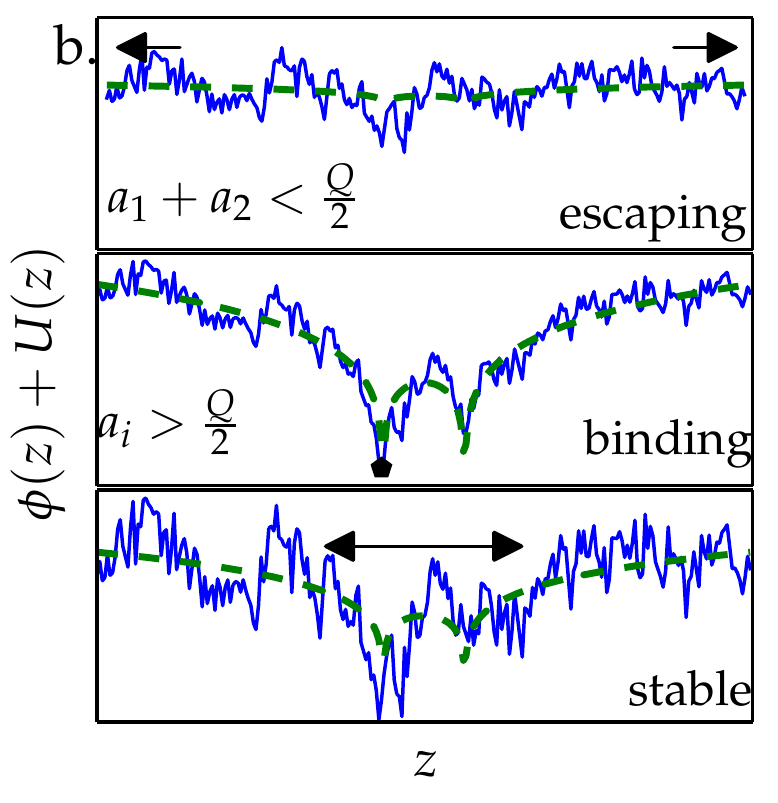}
\caption{(Colour online) a. Colour plot of a 2D GFF \eqref{eq:gff} sample plus the log confining potential $U(z)$ \eqref{eq:defUsimple} with $a_{1,2} = .8, .6$. The two singularities $z=0,1$ are indicated by dots. The domain has lattice spacing $\epsilon = 2^{-5}$ and size $R = 8$, with periodic boundary condition.  b. \red{Top: When eq. \eqref{eq:selberg1infty} is violated ($a_{1,2} = .1, Q = 2$), the particle is not confined, the $R \to \infty$ limit is ill-defined. Middle: When eq. \eqref{eq:selberg1} is violated  ($a_{1,2} = 2, Q = 2$), the particle is trapped and the Gibbs measure becomes a $\delta$ peak as $\epsilon \to 0$. Bottom: When both eq. \eqref{eq:selberg1} and \ref{eq:selberg1infty} are met $(a_{1,2} = .8, .6, Q = 2)$, the extent of the central region is stable as $R \to \infty, \epsilon \to0.$}} \label{fig:photo}
\end{figure}
Ideas of relating the Gibbs measure statistics in the 2D GFF and the $c\geq 25$-LFT go back to \cite{kogan96prelocalised,carpentier2001glass} (see \cite{david1992intrinsic} for earlier work on LFT--disordered system connections). Links were found between LFT features (scaling dimension, $c=25$ barrier) and disordered-system phenomena (multifractal exponents, freezing, respectively). However, as pointed out in \cite{carpentier2001glass}, these ideas were not fully exploited, because the asymptotic behaviour of the LFT field is subtle to implement in the statistical model under consideration. 
This Letter reopens the problem using more powerful methods, based on recent
progresses in LFT and in understanding of log-REM freezing transitions. 
Adding a logarithmic confining potential to the 2D GFF allows to establish an exact correspondence between the disorder averaged Gibbs measure in 2D and the LFT $4$-point function.
When carried through the freezing transition, this result leads to predictions for the probability distribution of the positions of the extrema in 2D and also extends to curved surfaces and higher order Gibbs measure correlations. \red{More generally, we use the short-distance behaviour of LFT correlators to give predictions that go beyond the previous setup and apply to \textit{all} log-REMs. This is possible thanks to the well-known dimension independence (universality) of many properties of logREMs \cite{carpentier2001glass,bolthausen2001} and mappings between them}. In particular, our results extend to arbitrary temperature a recent work of Derrida and Mottishaw \cite{derrida16kppfinitesize} on the directed polymer on a Cayley tree. The above outline is illustrated in Fig. \ref{fig:outline}

\textit{Set up}-- Our central object is the normalized Gibbs measure of a particle on the plane:
\begin{align}
& p_\beta(z) \defeq  \frac{1}{Z} e^{-\beta (\phi(z) + U(z))}  \,,\, z \in \C \,, \label{eq:boltzmann} \\ 
& Z \defeq \int_{\C}  e^{-\beta (\phi(z) + U(z))}  \dif^2 z \,.\label{eq:Z}
\end{align}
Here, $Z$ is the canonical partition function  at temperature $1 / \beta$, $U(z)$ is a confining potential defined as the sum of two logarithms:
\begin{equation}
U(z) \defeq 4 a_1 \ln |z| + 4 a_2 \ln |z - 1| \,,a_1, a_2 > 0  \label{eq:defUsimple}
\end{equation} 
and $\phi(z)$ is the 2D GFF. The latter is well-defined only in a finite geometry of size $R$ with a lattice spacing $\epsilon$. In the regime $\epsilon\ll\abs{z-w}\ll R$, the covariance is
\begin{equation}
\overline{\phi(z) \phi(w)} = 4 \ln (R/\abs{z-w}) \,,\,  \label{eq:gff}
\end{equation}
supplemented by $\overline{\phi(z)^2} = 4\ln(R/\epsilon)$ and $\overline{\phi(z)} = 0$. Figure \ref{fig:photo} shows a simulation of $\phi+U$. To prepare for the field theory connection below, we now discuss the $\epsilon\to0, R\to\infty$,  \textit{thermodynamic} limit of the model. For later convenience, the zero mode, immaterial for the Gibbs measure, is adjusted to vanish, \textit{i.e.} $\int \phi(z) \, \dif^2 z = 0$ for each realisation.

If one sets $U(z)=0$, the model belongs to the class of standard log-REMs, for which the mean free energy is universal (modulo an $O(1)$ correction) and displays a freezing transition at $\beta=1$ \cite{carpentier2001glass,derrida1988polymers,fyodorov08rem}:
\begin{eqnarray}
&\overline{F} = -Q \ln M + \eta \ln\ln M + O(1) \,,\, M = (R/\epsilon)^2 \,, \label{eq:F} \\
&Q = b + b^{-1} \,,\, b = \min (1, \beta) \,. \label{eq:defQ}
\end{eqnarray}
\red{Here, $ \eta \ln \ln M$ is the universal sub-leading correction \cite{bramson1983memoirs,derrida1988polymers,carpentier2001glass}. In the $\beta<1$ phase, it is absent ($\eta = 0$). At the critical point $\beta=1$, the sub-leading term appears with $\eta = \frac{1}{2}$. In the glassy phase $\beta > 1$, the leading term $-2\ln M$ displays freezing, and the correction coefficient becomes $\eta = \frac{3}{2}$.} Note that the leading behaviours are shared by the \textit{uncorrelated} Random Energy Model (REM) \cite{derrida1980random}, the first signature of the log-correlated universality being the sub-leading term $\frac{3}{2}\ln \ln M$ \cite{rosso12counting,fyodorov2015high}. 

When $U(z)$ is turned on, eq. \eqref{eq:F} may not persist. Indeed, when a log-singularity of $U(z)$ (say at $z=0$) is too deep, there can be a \textit{binding transition} \cite{carpentier2001glass,fyodorov2009statistical} dominating the free energy \red{(see Fig. \ref{fig:photo}. b/middle)}. This happens when the energy at its bottom is
$4a_{1} \ln \epsilon \ll \overline{F}$ as $\epsilon \to 0$, \textit{i.e.} when $a_1 > Q / 2$. This work excludes such bound phases, \red{in which the Gibbs measure is a trivial $\delta$}, by requiring
\begin{equation}  a_{1}, a_2  <  Q / 2 \,. \label{eq:selberg1} \end{equation} 
Moreover, the potential must also confine the particle at $z \sim O(1)$ in the $R \to +\infty$ limit; \red{otherwise $p_\beta$ would be non--normalizable in that limit (see Fig. \ref{fig:photo} b./top)}. Thus, we require $\overline{F} + U(R) \to +\infty$ as $R \to +\infty$, or
\begin{equation}
 a_1  +  a_2 > Q / 2 \,. \label{eq:selberg1infty}
\end{equation}
When \eqref{eq:selberg1} and \eqref{eq:selberg1infty} are satisfied, $p_\beta (z)$ has a well-defined non-trivial limit (in law) as $\epsilon \to0, R\to \infty$ \red{(see Fig. \ref{fig:photo} b./bottom)}. Thus, adding a confining potential is sufficient to make the position problem well-posed. By contrast, the free energy distribution is dominated by long-wave-length fluctuations of $\phi$ and suffers from an $R \to \infty$ divergence, whose proper subtraction is an open question (see however discussions in 1D \cite{fyodorov2010freezing, fyodorov2009statistical, cao15gff}).

\textit{Connection to LFT in $\beta<1$ phase}-- Let us first introduce some notations. Let  $\left<\prod_{i=1}^{n}\vertex_{a_i}(z_i)\right>_{b}$ be the \red{Euclidean} $n$-point correlation function of the LFT defined on the complex plane plus a point at $\infty$, $\C \cup \{\infty\}$, and with central charge $c=1+6Q^2, \red{Q = b + b^{-1}}$. The field $\vertex_{a}(z)$ is a primary field with scaling dimension $\Delta_{a}=a(Q-a)$\cite{zamolodchikov2007lectures,SM}. We first demonstrate the connection between the Gibbs measure statistics and LFT on the simplest example. We claim:
\begin{align}
& \overline{p_\beta(z)} \stackrel{\beta < 1}\propto \left\langle  \vertex_{a_1} (0)  \vertex_{a_2} (1) \vertex_b(z) \vertex_{a_3} (\infty)  \right\rangle_{b} \label{eq:mainhighT}
\end{align}
where $a_3=Q-a_1-a_2$ \footnote{Since $\int d^2 z \, \overline{p_{\beta}(z)} = 1$, the proportionality constant can be evaluated once the LFT correlation is known.}. 

 In order to show the above identity, we will use the LFT functional integral representation. This is defined, on any closed surface $\Sigma$, from the action $\mathcal{S}_b$:
\begin{equation}
\mathcal{S}_b = \int_{\Sigma} \left[ \frac{1}{16\pi} (\nabla \varphi)^2 - \frac1{8\pi}Q \hat{R} \varphi + \mu e^{-b\varphi}  \right] \dif A  \,,\label{eq:Sliouville} 
\end{equation}
where $\mu$ is the coupling constant, $\hat{R}$ is the Ricci curvature and $\dif A$ the surface element. Note that in our case, the  surface  $\Sigma=\C \cup \{\infty\}$ has the topology of a sphere with the curvature  concentrated at $\infty$ and vanishing elsewhere: $\hat{R}(z) =  8\pi \delta^2(z-\infty), \dif A = \dif^2 z$. In this representation, the primary fields are exponential fields, $\vertex_a(w) \leadsto e^{-a\varphi(w)}$, also called \textit{vertex operators}. The $4$-point correlation function in \eqref{eq:mainhighT} can be written as:
\begin{align}
K_4 \defeq & \int \mathcal{D}\varphi\, e^{-\mathcal{S}_b -b \varphi(z) -a_1\varphi(0)-a_2 \varphi(1) - a_3 \varphi(\infty)} \,,\, \label{eq:fcnint}
\end{align}
where we noted $K_4\defeq \left\langle  \vertex_{a_1} (0)  \vertex_{a_2} (1) \vertex_b(z) \vertex_{a_3} (\infty)  \right\rangle_{b}$ for better readability. To derive \eqref{eq:mainhighT}, we recall \red{that the Liouville field is decomposed into a zero mode and a fluctuating part, $\varphi(z)=\zmode + \tilde{\varphi}(z)$, where $\zmode$ is the zero mode \cite{goulian1991correlation} (see \cite{david2014liouville} for recent rigorous work in a related setting). Accordingly, the functional integral is written as $\int \mathcal{D}\varphi = \int_\R \dif \zmode \int \mathcal{D}\tilde{\varphi}$. Once we performed the integration over $\zmode$, the one over $\tilde{\varphi}$ can be written as an expectation over the 2D GFF \textit{without} zero mode, \textit{i.e.}, over $\phi$ defined in eq. \eqref{eq:gff}, that is, for any observable $\mathcal{O}$, we have $\int \mathcal{D}\tilde{\varphi} e^{- \int \frac{1}{16\pi} (\nabla \tilde{\varphi})^2 \dif^2 z } \mathcal{O}[\tilde{\varphi}]  = \overline{\mathcal{O}[\phi]}$. With these considerations one can obtain} 
\begin{align} 
\mu b K_4 &=&  \overline{e^{- a_1\phi(0)-a_2 \phi(1) + (a_1+a_2)\phi(\infty)-b\phi(z)} / Z_0} \,,\label{eq:step2}
\end{align}
where $Z_0 = \int_{\C} e^{-b\phi(z)} \dif^2 z.$ The choice of $a_3$ in \eqref{eq:mainhighT} is crucial for the apparition of $Z_0^{-1}$. Then, a complete-the-square trick allows to identify the average in \eqref{eq:step2} to $\overline{p_\beta (z)}$, leading to \eqref{eq:mainhighT} (see \cite{SM} for details).

The above steps generalize easily to the multi-point correlations of \red{powers of the Gibbs measure $p_\beta^{q_i}(z_i) = (p_\beta(z_i))^{q_i}$, $q_i \geq 0$,} with $U(z)=\sum_{j=1}^k 4 a_j \ln\abs{z-w_j}$ such that $\forall a_j < Q / 2$ and $a_{k+1} \defeq Q - \sum_{j=1}^k a_j < Q / 2$ (compare to \eqref{eq:selberg1} and \eqref{eq:selberg1infty}). The result is stated as 
\begin{align}
 & \overline{\prod_{i=1}^n p_\beta^{q_i}(z_i)}  
 \stackrel{\beta<1}\propto  
 \left< \prod_{j=1}^{k+1} \vertex_{a_j} (w_j) \prod_{i=1}^n \vertex_{\beta q_i} (z_i) \right>_{b} \label{eq:general}
\end{align}
where $w_{k+1}=\infty$ and $\forall q_i < Q / (2 \beta)$ \cite{SM}. Moreover, \eqref{eq:general} holds, in general closed surfaces \cite{SM}. While mapping Gibbs measure correlations onto LFT correlations on a sphere requires a potential with $\geq 3$ singular points (\textit{e.g.}, $0,1$ and $\infty$ for \eqref{eq:defUsimple}), on a torus the potential is unnecessary. \red{In general, the sum of the charges must be equal to $Q \chi /2 $, where $\chi$ is the Euler characteristics of the surface ($\chi = 2$ for the sphere and $0$ the torus) \cite{SM}.}

We now use known properties of LFT to obtain new results for our log-REM model, and beyond.

\textit{$\beta>1$ phase}--The $4$-point function in \eqref{eq:mainhighT} is invariant under the transform $b\to 1/b$ \cite{zamolodchikov1996conformal}. Hence, from the \textit{freezing-duality conjecture} \cite{fyodorov2009statistical,fyodorov2016interval}, we expect that $\overline{p_{\beta > 1}}= \overline{p_1}$ \textit{freezes}, so the prediction \eqref{eq:mainhighT} \textit{still holds} thanks to the notation $b = \min(1,\beta)$; this can be also shown by replica symmetry breaking (RSB) \cite{fyodorov2010freezing,cao16order}. Taking the $\beta\rightarrow \infty$ limit gives the position distribution of the minimum of $\phi(z) + U(z)$. Note that the freezing of $\overline{p_\beta}$ does \textit{not} imply that of $p_\beta$, as revealed by its multi-point correlations. Indeed, $p_{\beta > 1}$ develops $\delta$ peaks, which are absent when $\beta < 1$, and which give rise to a $\delta$ contact singularity in $2$-point correlations of $p_{\beta>1}$. For example, an RSB calculation as in \cite{fyodorov2010freezing,cao16order} gives 
\begin{equation} \overline{p_\beta(z_1)p_\beta(z_2)} = (1-T) \delta_{1,2}\, \overline{p_1(z_1)} + T \, \overline{p_1(z_1)p_1(z_2)} \,,\label{eq:onestep} 
\end{equation} 
where $\delta_{1,2} = \delta(z_1 - z_2)$ and $T = 1/\beta<1$. We will further apply and discuss this result below, see eq. \eqref{eq:overlap}.

At $\beta\to\infty$, the positions of the deepest minima of the 2D GFF can be also studied by RSB \cite{cao16order} (see also some rigorous results \cite{biskup2016full}). That allows us to show, for instance, that the joint probability distribution of the first and second minima positions ($\xi_{1,2}$) is (see \cite{SM}):
\begin{equation} 
P(\xi_1,\xi_2) = c_0 \delta(\xi_1-\xi_2) \overline{p_1(\xi_1)}  + (1-c_0)\overline{p_1(\xi_1) p_1(\xi_2)} \label{eq:min12}
\end{equation}  
and thus also relates to LFT using (\ref{eq:general}). Here $c_0=1-\overline{g}$ is the probability that the two lowest minima belong to the same ``cluster'' and $g$ is the energy gap between them, which depends on model-specific details at the $\sim\epsilon$ scale.  

\begin{figure}
\begin{center}
\includegraphics[scale=.37]{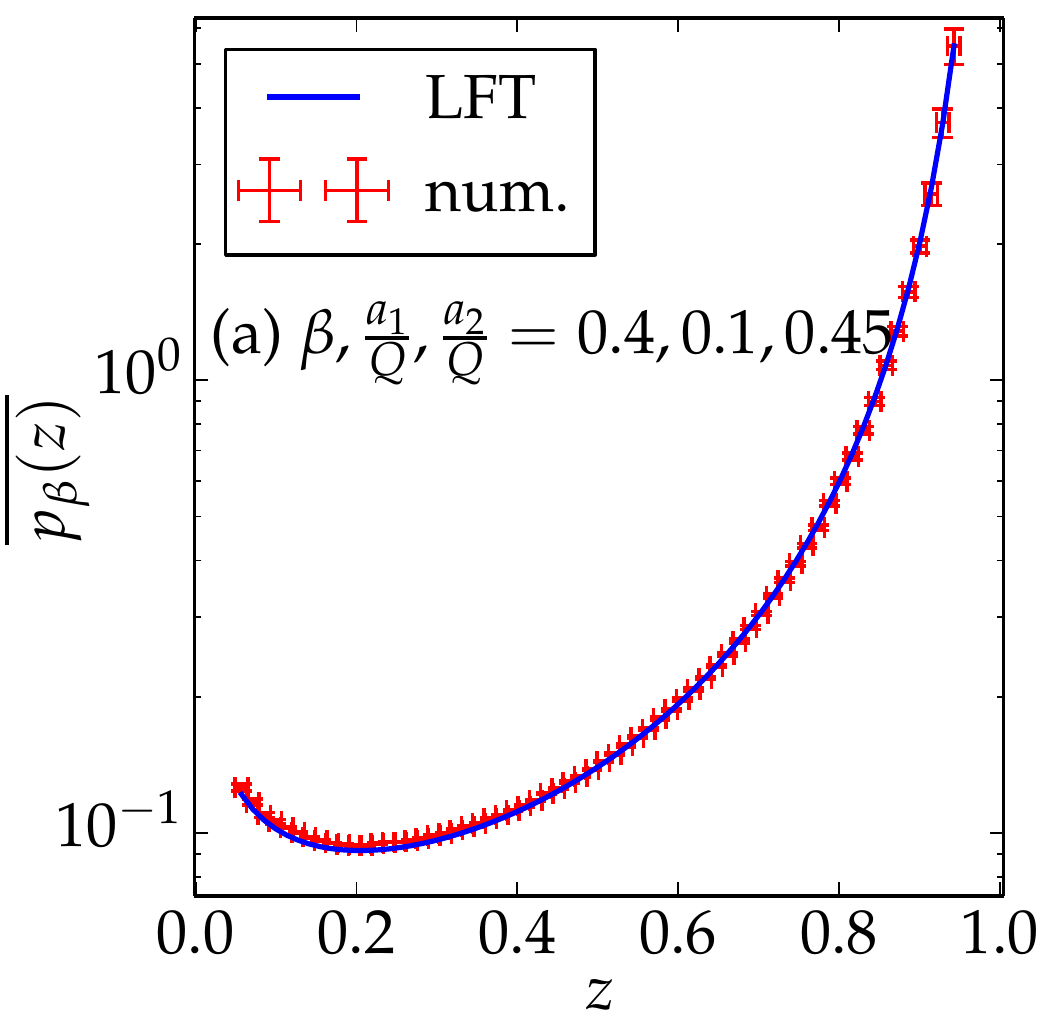}
\includegraphics[scale=.37]{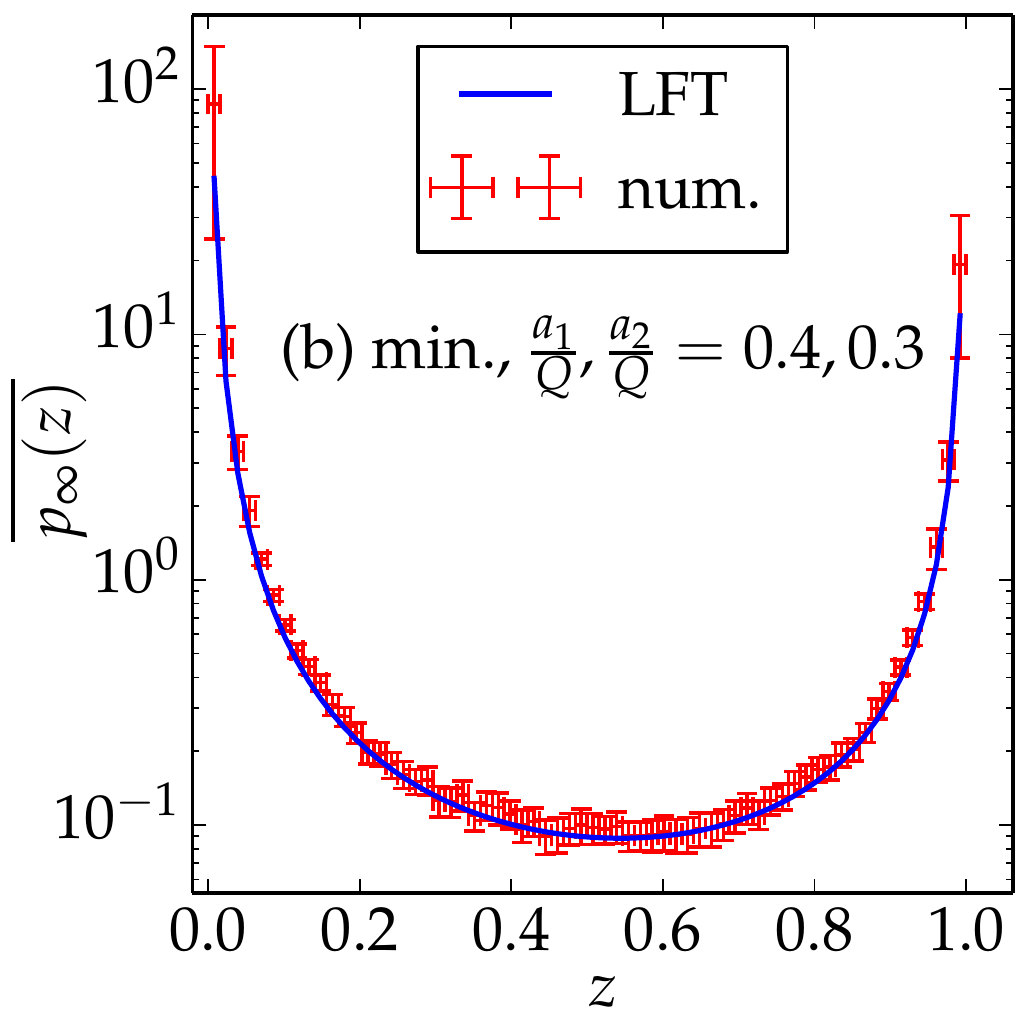}
\end{center}
\caption{(Color online) Test of \eqref{eq:mainhighT} on the segment $z\in [0,1]$. (a) High-$T$ regime ($\beta=.4$). (b) Minimum position distribution versus LFT with $b=1$. Numerical parameters: $L = 2^{12}, \epsilon = 2^{-9}$, $5 \times 10^6$ independent samples.} \label{fig:tests}
\end{figure}
\textit{Numerical test}-- The LFT $4$-point functions are exactly calculated by the conformal bootstrap  \cite{zamolodchikov1996conformal,SM}, implemented by the code-base \cite{codebootstrap}, extended to take into account the \textit{discrete terms} (they have important consequences, see below). The LHS of \eqref{eq:mainhighT} is measured on extensive simulations of discrete 2D GFF, as shown in Figure \ref{fig:photo}. The results validate unambiguously the predictions, see Figure \ref{fig:tests}. Now that the advocated relation has been confirmed in a particular setting, the next goal is to extract more universal physical consequences from LFT. 

\textit{Liouville OPE}--As can be seen in Fig. \ref{fig:tests}, $\overline{p_\beta(z)}$ diverges as $z$ comes near a log singularity of the potential $U(z)$, say as $z\to 0$ where $U(z) \approx 4 a_1 \ln \abs{z}$. This asymptotic behaviour depends only on $\beta$ and $a_1$, and can be obtained from an \textit{operator product expansion} (OPE) $\vertex_{\alpha}(0)\vertex_{\alpha'}(z)$ \cite{ribault2014conformal}. Such OPE's have been obtained by conformal bootstrap \cite{SM} and read as follows:
\begin{align}
& \langle \vertex_{a}(0)\vertex_{a'}(z) \dots \rangle_b  \stackrel{z\to 0}\sim  \begin{cases}
 \abs{z}^{-2\delta_0} ,\,  a''\defeq a+a' < \frac{Q}{2} \,, \\
 \abs{z}^{-2\delta_0}\ln^{-\frac{1}{2}}\abs{1/z},\, a'' =\frac{Q}{2} \,, \\
 \abs{z}^{-2\delta_1}\ln^{-\frac{3}{2}}\abs{1/z} ,\,  a''>\frac{Q}{2}  \,,
 \end{cases} \nonumber \\
&\delta_0 =2aa',\, \delta_1 = \Delta_{a} + \Delta_{a'} - \Delta_{\frac{Q}{2}},\, \Delta_a = a(Q - a) \label{eq:ope}\, 
\end{align} 
These asymptotic behaviours hold for generic LFT correlations, as long as the distance $\abs{z}$ is much smaller than that to the other operators (as well as $R$). Note moreover that field theory predictions break down when $\abs{z} \sim \epsilon$. To obtain the divergence of $\overline{p_\beta(z\to 0)}$ shown in Fig. \ref{fig:tests} from \eqref{eq:mainhighT}, we must set $a = a_1$ and $a' = b$ in \eqref{eq:ope}.

The abrupt behaviour change as the parameters cross the line $a + a' = Q / 2$ comes from a peculiar feature of LFT and corresponds to the presence/absence of the \textit{discrete terms} \cite{teschner01LFT,teschner2015supersymmetric,belavin2006integrals,aleshkin2016construction} (see also \cite{ribault2014conformal}, Ex. 3.3 and \cite{SM}). To discuss the physical consequences of this feature, we consider two independent thermal particles in one realisation, and the disorder-averaged joint position distribution $\overline{p_\beta(w)p_\beta(z+w)}$. If $w$ is fixed far from singularities, and $z \to 0$, the asymptotic dependence on $z$ is given by eq. \eqref{eq:ope} (with $a =  a' = b$), independently of the other details. In particular, combining with \eqref{eq:min12} gives the following asymptotics of the first-second minima positions distribution:
\begin{equation}
P(\xi_1, \xi_2) \sim  \abs{\xi_1 - \xi_2}^{-2} \ln^{-\frac{3}{2}}\abs{1/(\xi_1 - \xi_2)} 
\end{equation} 
which holds for $1 \gg \abs{\xi_1 - \xi_2} \gg \epsilon$ (while the $\delta$ in \eqref{eq:min12} takes over as $\abs{z}\sim \epsilon$). 

\textit{Beyond 2D}--The robustness of the above results suggests their generalization beyond 2D GFF models to general log-REMs, such as the directed polymer on disordered Cayley tree model \cite{derrida1988polymers}. \red{This is the best studied logREM, due to its relevance in classical (\textit{e.g.}, Kardar--Parisi--Zhang class \cite{Halpin-Healy2015KPZ}) and quantum (\textit{e.g.} Anderson transition \cite{evers2008anderson}) disordered systems.}  It is defined on a Cayley tree (see Fig. 3 of \cite{SM}) of depth $t$ and branching number $\kappa$ ($\kappa = e$ for Branching Brownian Motion). Each edge has an independent Gaussian random energy of zero mean and variance $2 \ln \kappa$ (so that freezing occurs at $\beta=1$). A directed polymer (DP) is a simple path from the root to some leaf, and its energy is the sum of the edge-energies. Then, the energy of all the DP's $\phi_1, \dots, \phi_{M}, M = \kappa^t$ are centered Gaussian with covariance
\begin{equation} \overline{\phi_i \phi_j} = 2 \hat{\overlap}_{ij} \ln \kappa \,, \label{eq:covtree} \end{equation} 
where $\hat{\overlap}_{ij} \in [0,t]$ is the common length of $i$ and $j$. An interesting question is the distribution $P(\hat{\overlap})$ of the common length of two independent thermal DP's drawn from a single Gibbs measure $p_i \propto e^{-\beta \phi_i}$ for $t \to \infty$. This quantity is \textit{different} from the more studied distribution of the \textit{overlap} $\overlap=\hat{\overlap}/t$  for $t\to\infty$  \footnote{Its limit law is $\min(T,1)\delta(\overlap) + (1-\min(T,1))\delta(1-\overlap)$ \cite{derrida1988polymers,arguin2011genealogy} here, but is more involved in other situations of current interest \cite{franz2016mean}}. Our results correspond to the leading finite-$t$ correction of the latter near $ \overlap= 0$.

To calculate $P(\hat{\overlap})$, we compare positions/distance in 2D to DP's/common length on the tree, by matching the respective covariances \eqref{eq:gff} and \eqref{eq:covtree}. This gives $\abs{z} = r = \kappa^{-\hat{\overlap}/2} \in [\kappa^{-\frac{t}{2}}=\epsilon, 1=R]$, leading to the transformation $P(\hat{\overlap}) \dif \hat{\overlap} = \overline{p_\beta(w)p_\beta(w+r)} 2\pi r \dif r $. Then applying \eqref{eq:ope} leads to  (see \cite{SM}, eq. (35))
\begin{equation} 
P(\hat{\overlap})  \sim 
\begin{cases}
\kappa^{(2 \beta^2 - 1)\hat{\overlap}} \,,&  \beta < 3^{-\frac{1}{2}},  \\
\kappa^{- \hat{\overlap} /3} \hat{\overlap}^{-\frac{1}{2}}\,, & \beta = 3^{-\frac{1}{2}}, \\
\kappa^{- (\beta - \beta^{-1})^2 \hat{\overlap}/4} \, \hat{\overlap}^{-\frac{3}{2}} \,,&  \beta \in (3^{-\frac{1}{2}},1) \, \\
 \hat{\overlap}^{-\frac{3}{2}} \beta^{-1},  & \beta \geq 1 \,, \hat{\overlap} \ll t\,.
\end{cases}  \label{eq:overlap}
\end{equation} 
For the $\beta\geq1$ case we used also the RSB result \eqref{eq:onestep}, which can be interpreted as follows: with probability $T=1/\beta$, the common length $\hat{\overlap}$ remains finite, and the Liouville OPE applies; while with probability $1-T$, $\hat{\overlap} \sim O(t)$. In the 2D context, the latter case corresponds to two particles \textit{frozen} at a distance $\sim \epsilon$. The field theory results are valid only in the continuum regime $ r \gg \epsilon$, which corresponds to $ \hat{\overlap} \ll t$ for the DP model. For this reason, our $\beta \geq 1$ result matches the exact solution of \cite{derrida16kppfinitesize} when $\overlap = \hat{\overlap}/t \ll 1$, and loses validity at $\overlap\to 1$ ($\hat{\overlap} \to t$). 

The results for the $\beta < 1$ phase are new, and display, in the high-$T$ phase, log-corrections typical of the freezing transition. As will be reported in up-coming work, they are universal signatures of the \textit{termination point transition} (called ``\textit{pre-freezing}'' in \cite{fyodorov2009pre}) in log-REMs. This transition manifests itself in the \textit{annealed} average of inverse partition ratio $P_q \defeq \sum_{i=1}^M (e^{-\beta \phi_i} / Z)^q$, $q > 0$. Indeed, one can show
\begin{align} - \ln \overline{P_q} \stackrel{\beta<1}\sim  \begin{cases}  
\tau(q) \ln M & q \beta  < \frac{Q}{2}  \\
\tau(q) \ln M + \frac{1}{2} \ln \ln M  & q \beta  =  \frac{Q}{2} \\
\tau(q) \ln M + \frac{3}{2} \ln \ln M  & q \beta  > \frac{Q}{2}  \\
\end{cases} \,, \label{eq:Pq}
 \end{align}
 where $\tau(q) = \Delta_{\min(q\beta, Q/2)} - 1$ (see \eqref{eq:ope}). Note that for the uncorrelated REM \cite{fyodorov2009pre}, we would have the same $\tau(q)$ but $\frac{1}{2}\ln\ln M$ correction when $q\beta > Q/2$ \footnote{Log-log corrections for the {\it quenched} average (\textit{typical} samples) were known, Sec. VI.B in \cite{carpentier2001glass}; their occurrence in the annealed ensemble considered here, which probes also \textit{rare} samples, is new.}. In LFT, the latter phase is where $p_\beta^q$ can no longer be represented by $\vertex_{q \beta}$ as it would violate a Seiberg bound (\eqref{eq:general} and \cite{SM}).
 
\textit{Conclusion}--We related $c\geq25$ LFT to the Gibbs measure of 2D GFF plus log potential, and found indications that LFT may describe universal features of general log-REMs. \red{We mention two exciting perspectives. The first is extending the mapping to logREMs with \textit{imaginary temperature}  ($b\to \im b$), where relations to $c\leq1$ conformal field theories, $\beta$-Random Matrix Ensembles and 2d log-gases are natural to expect. The other concerns the glassy phase $\beta > 1$, in which LFT must be supplemented by RSB/freezing-duality conjecture in order to make correct predictions. However, the termination point transition predicted by LFT alone resembles strikingly the freezing transition. This points to the intriguing question: Does the glassy phase have a field theory description?}

\begin{acknowledgements}
We thank K. Aleshkin, Vl. Belavin, Y. V. Fyodorov, A. Morozov, S. Ribault, R. Rhodes and V. Vargas, for enlightening discussions; the Organisers of the workshop ``\textit{Quantum integrable systems, conformal field theories and stochastic processes}'', during which this work was initiated; l'Institut d'\'Etudes Scientifiques de Cargèse for their hospitality, and Corsica for her beautiful land, sea and people. XC acknowledges the support of Capital Fund Management Paris and LPTMS.
 \end{acknowledgements}

 \numberwithin{equation}{subsection}
 
 \onecolumngrid
 
\pagebreak

\section*{Supplementary Material}
 The Supplementary Material is organized as follows. First we treat the technical points in the derivation of the main result: the geometry of the flat plane plus a point at infinity and the completing-the-square trick. Then we provide more details on the general result: Seiberg bounds, and details on the result on general surfaces. Then we give a brief introduction to the conformal bootstrap approach to LFT $4$-point functions and OPE's, focusing on the the discrete terms. Finally we give provide technical backgrounds from the theory of logREMs. 
 
 \subsection{Main result}
 \subsubsection{The infinite plane plus a point}\label{sec:flat}
 The infinite plane plus a point is topologically identical to the round sphere, but its geometry resembles more the flat Euclidean plane, except that an infinite point is added to it. Moreover that point is singular. To describe this more properly, one may take two complex planes, one parametrized by $z$ and the other $w$, and glue them together by the mapping $z = 1 / w$, so the infinite point $z \to \infty$ in one plane corresponds to $w \to 0$ in the other. 
 The line element (metric) on the $z$-plane is defined as $\dif s = \abs{\dif z}$, and transforms to $\abs{w^{-2} \dif w}$ in the other coordinate, so the area element writes as $\dif A = \dif^2 z = \abs{w^{-4}} \dif^2 w$. The Ricci curvature is a delta peak $ \hat{R}(w) = 8 \pi \abs{w^4} \delta(w) = 8 \pi \delta(z-\infty)$, so that $\int_\C \hat{R}(w) \dif A = 4 \pi \chi$ with Euler characteristics $\chi = 2$, satisfying the Gauss-Bonnet theorem, which states that for any closed surface $\Sigma$ with Euler characteristics $\chi$, curvature $\hat{R}$ and surface element $\dif A$, 
 \begin{equation}
  \int_{\Sigma}  \hat{R}   \dif A   =  4 \pi \chi \,.  \label{eq:gaussbonnet}
 \end{equation}

  \subsubsection{Complete the square}\label{sec:square}
 We recall the complete-the-square trick (also known as Girsanov transform \cite{girsanov1960transforming}, see \cite{david2014liouville} for rigorous treatment in similar context) applied to the following average over the 2D GFF $\phi(z)$:
 \begin{align} 
 & \overline{  \exp\left[ - \int u(z) \phi(z) \dif^2 z \right] \mathcal{O}[\phi]  } = \int \mathcal{D}\phi  
 \exp\left[ \int \left( \frac{1}{16\pi} \phi\Delta\phi - u \phi \right) \dif^2 z  \right] \mathcal{O}[\phi]  \nonumber \\
  = &  C \int \mathcal{D}\phi   \exp\left[ \int \left( \frac{1}{16\pi} (\phi-U) \Delta (\phi-U) \right) \dif^2 z  \right] \mathcal{O}[\phi] 
   \nonumber \\
  = & C  \int \mathcal{D}\phi   \exp\left[ \int \left( \frac{1}{16\pi} \phi \Delta \phi \right) \dif^2 z  \right] \mathcal{O}[\phi + U]
  =  \overline{\mathcal{O}[\phi + U]} \,, \\
 \text{where }&  U = 8 \pi \Delta^{-1} u  \,,\, C =  \exp\left( - \frac{1}{16 \pi} \int U \Delta U \dif^2 z \right) \,, \label{eq:complete}
 \end{align}
 and $\mathcal{O}[\phi]$ is any observable (functional) of $\phi$. The Poisson equation $U = 8\pi \Delta^{-1} u$ has a solution on a closed surface (here it is the flat background) if and only if $u(z)$ fulfils the \textit{neutrality condition}
\begin{equation} \int u(z) \dif^2 z = 0  \,. \label{eq:neutrality}\end{equation}
When the solution exists, it is unique up to a global shift. In our application (eq. \eqref{eq:step2} in the main text), 
\begin{itemize}
\item $u(z) = a_1 \delta(z) + a_2 \delta (z-1) - (a_1 + a_2) \delta(z-\infty)$. It satisfies the neutrality condition \eqref{eq:neutrality}, and a solution is given log confining potential $U(z) = 4 a_1 \ln\abs{z} + 4a_2 \ln \abs{z-1}.$ 
\item $\mathcal{O}[\phi] = e^{-b\phi(z)} Z_0^{-1}$,  where $Z_0 = \int e^{-b\phi(z)} \dif^2 z$, so 
$\mathcal{O}[\phi + U] = e^{-b(\phi(z) + U(z))} Z^{-1} = p_\beta(z)$ by definition.
\end{itemize}  
In summary, the complete-the-square trick \eqref{eq:complete} shows that $\eqref{eq:step2}$ in the main text is equal to $C \overline{p_\beta(z)}$, \textit{i.e.}, they are equal up to a $z$-independent factor. 

 \subsection{General result}
 \subsubsection{Seiberg bounds}\label{sec:seiberg}
 We show here that the LFT correlation functions used in the main text
 \begin{equation}
\left< \prod_{j=1}^{k+1} \vertex_{a_j}(w_j) \prod_{i=1}^ n \vertex_{q_i \beta}(z_i) \right>_b \label{eq:corre}
 \end{equation}
  satisfy the Seiberg bounds \cite{seiberg1990notes}, which are the sufficient and necessary condition for their rigorous probabilist construction \cite{david2014liouville}. There are two types of Seiberg bounds:
 \begin{itemize}
 \item[\texttt{i.}] All the \textit{charges} $a_1, \dots, a_{k+1}$ and $q_1 \beta, \dots, q_n \beta$ must be $< Q / 2$. For $a_1, \dots, a_k$, this is the same as the no-binding condition \eqref{eq:selberg1} in the main text. For $a_{k+1}$, it is the confinement condition \eqref{eq:selberg1infty} in the main text. Finally the bounds $q_i \beta < Q / 2$ are the ones required in the main text. The line $q  \beta = Q / 2$ is also that of the termination point transition.
 \item[\texttt{ii.}] The sum of all the charges is larger than $Q$. In our case, the sum is $\sum_{i=1}^n b q_i + \sum_{j=1}^{k+1} \alpha_j = \sum_{i=1}^n b q_i + Q$. So as long as $n > 0$ and $q_i > 0$, this bound is satisfied. 
   \end{itemize}
   
 \subsubsection{General curved surfaces}
 We give the precise statements of the LFT--Gibbs measure connection in a general closed Riemannian surface $\Sigma$ parametrized by a local complex coordinate $z$, and with area element $\dif A(z)$ and line element $\dif s$. The statistical physics model is generalized as follows:
 \begin{itemize}
 \item[-] The partition function and Gibbs measure generalize naturally as
 \begin{equation}
 Z = \int_\Sigma e^{-\beta (\phi(z) + U(z))} \, \dif A(z) \,,\, p_\beta(z) = \frac1Z e^{-\beta(\phi(z) + U(z)}  \,.
 \end{equation}
 \item[-] The 2D GFF $\phi(z)$ is defined on $\Sigma$ in the standard way, \textit{i.e.}, by the covariance kernel which is the Green function of the Laplace equation on the closed surface
 \begin{equation}
\overline{\phi(z) \phi(w)} = K(z,w) \,,\, \Delta_z K(z,w) = 8 \pi \left(V^{-1} - \delta_{z,w}\right)  \,, \int_\Sigma K(z,w)  \dif A(z) =0 \,. \label{eq:generalgff}\end{equation}  
Here, $V = \int_\Sigma \dif A$ is the surface area, $\Delta_z$ is the Laplace-Beltrami operator on the surface $\Delta$, here applied on the $z$ variable, and $\delta_{z,w}$ is the Dirac delta with respect to the area form, \textit{i.e.}, satisfying $\int_\Omega \delta_{z,w} \dif A(z)  = 1$ if $w \in \Omega$ and $0$ otherwise. The last condition in \eqref{eq:generalgff} fixes the zero mode of $\phi$ to vanish. The UV regularization is done by a lattice spacing $\epsilon$ that is uniform in the surface length unit (\textit{not} in local coordinate unit), so that $\overline{\phi(z)^2} = -4 \ln \epsilon + c_1$ with $c_1$ a $z$-independent constant.
\item[-] To define the potential $U(z)$, we need to specify a sequence of points $w_1, \dots, w_{\ell}  \in \Sigma$ and of \textit{charges} $a_1, \dots, a_{\ell} \in \R$ (note that $\ell = k+1$ in the main text). 
 $U(z)$ is a solution to the following Poisson equation: 
\begin{equation}
\Delta U(z) = 8\pi( a_1 \delta_{z,w_1} + \dots + a_{\ell} \delta_{z,w_\ell})  -Q\hat{R} \,. \label{eq:Ugeneral}
\end{equation} 
The charges should satisfy:
\begin{equation}
a_1, \dots, a_{\ell} < Q / 2 \,,\, a_1 + \dots + a_\ell = Q \chi / 2 \,, \label{eq:generalbound}
\end{equation}
where $\chi$ is the Euler characteristic of $\Sigma$. Note that according Gauss-Bonnet theorem (eq. \eqref{eq:gaussbonnet}) the second condition in \eqref{eq:generalbound} ensures that eq. \eqref{eq:Ugeneral} has a solution, which then is unique up to a global constant. 

If $\Sigma$ has the sphere topology, $\chi = 2$, \eqref{eq:generalbound} implies then that $\ell \geq 3$. So the setting in the main text is the minimum to compare to LFT. Now if $\Sigma$ is a torus, for which $ \chi = 0$, there can be no charges, \textit{i.e.}, we can set $\ell = 0$, $U(z) = 0$. Finally, for higher genus surface, $\chi < 0$, \textit{repulsive} log-singularities will be needed. 
 \end{itemize}
With these definitions, the derivation in the main text can be generalized to show the following:
\begin{equation}
\overline{p^{q_1}_\beta(z_1) \dots p^{q_n}_\beta(z_n)} \stackrel{\beta<1}\propto 
\left<  \prod_{i=1}^n \vertex_{\beta q_i} (z_i) \prod_{j=1}^{\ell} \vertex_{a_j}(w_j) \right>_{b,\Sigma} ,\, q_1, \dots, q_n   \in (0,Q/(2\beta)) \,, \label{eq:generalSM}
\end{equation}
which is just eq. \eqref{eq:general} in the main text, with $\ell = k+1$ and $w_{\ell}$ not necessarily equal to $\infty$.
On the left hand side, we consider correlations of \textit{fractional powers} of the Gibbs measure, $p_\beta^{q}(z) \defeq e^{-q\beta\phi(z)} Z^{-q}$; we stress that the temperature associated to the Gibbs measure and the partition function is \textit{not} multiplied by $q$. On the right hand side, the LFT correlation function is evaluated on the surface $\Sigma$, by the same functional integral formula \eqref{eq:fcnint} in the main text. When $\chi = 2$, LFT correlation functions on $\Sigma$ can be simply related to the flat plane ones (thus computed by the conformal bootstrap), as we recall in the following section. 

We sketch the derivation of \eqref{eq:generalSM} in two steps. The first step follows again \cite{goulian1991correlation}: from the functional integral definition of the RHS of \eqref{eq:generalSM}, one integrates the zero mode, and obtains
$$ \text{RHS of } \eqref{eq:generalSM} = \text{const. }  \overline{ Z_0^{-s/b} \prod_{i=1}^n e^{-q_i\beta \phi(z_i) }\prod_{j=1}^{\ell} e^{-a_j \phi(w_j) } \exp\left( \int_\Sigma \frac{Q \hat{R}\phi}{8\pi} \dif A \right) } \,,\, Z_0 = \int_\Sigma e^{-b\phi(z)} \dif A(z)  \,, $$ 
where $s = \sum_{i=1}^n q_i \beta  +\sum_{j=1}^\ell a_j - \chi Q / 2 =   \sum_{i=1}^n q_i \beta$ by \eqref{eq:generalbound}, and $\overline{[\dots]}$ denotes the expectation over the GFF on $\Sigma$ defined in \eqref{eq:generalgff}. The second step is a complete-the-square trick (analogous to that explained above) that absorbs the linear exponentials 
$$\prod_{j=1}^{\ell} e^{-a_j \phi(w_j) } \exp\left( \int_\Sigma \frac{Q \hat{R}\phi}{8\pi} \dif A \right) 
=  \exp\left(-\int_\Sigma \varphi(z) u(z) \dif A(z)\right) \,,\, u(z) =  \sum_{j=1}^\ell a_j \delta_{z , w_j} - Q \frac{\hat{R}}{8\pi} $$
into the potential $U(z)$ defined by \eqref{eq:Ugeneral}, which is just $\Delta U(z) = 8 \pi u(z)$. Note that two steps produce precisely the correct power of the normalization factors $Z_0$ (first step), and, finally, $Z$ (second step) at the denominator of the r.h.s. of \eqref{eq:generalSM}.
 
 \subsubsection{LFT on curved surfaces and conformal covariance}
 We recall how LFT correlation functions on different curved surfaces are related to each other, referring to \cite{zamolodchikov2007lectures} for more detailed introduction and \cite{david2014liouville} for modern rigorous treatment. For simplicity, we focus on the case where the surface has the topology of a sphere, and relate LFT correlations on it to those on the flat plane plus a point at infinity. In this case, we can assume that the surface is parametrized by $z \in \C $ (plus an infinity point) such that the line and surface elements are
  \begin{equation}
  \dif s = e^{\sigma(z)} \abs{\dif z} \,,\, \dif A = e^{2\sigma(z)} \dif^2 z \,.
  \end{equation}
  Now the LFT $n$-point function on this curved surface and that on the flat background are related by 
   \begin{equation}
   \left\langle  \prod_i \vertex_{\alpha_i} (z_i)  \right\rangle_{b,\text{curved}} = 
   \left\langle  \prod_i \vertex_{\alpha_i} (z_i)  \right\rangle_{b,\text{flat}} \prod_i e^{-2 \alpha_i(Q - \alpha_i) \sigma(z_i)}
    \times   \mathcal{N}_{\text{Weyl}}[\sigma]
   \end{equation}
    where $\mathcal{N}_{\text{Weyl}}[\sigma]$ is a global factor depending on the new geometry, known as the \textit{Weyl anomaly}. In some works, \textit{e.g.}, \cite{david2014liouville}, vertex operators are defined differently so as to absorb the scaling factors $e^{-2 \alpha_i(Q - \alpha_i) \sigma(z_i)}$. \vspace{.3cm}
    
  Closely related to above is the \textit{conformal covariance} property of the LFT. In our context it gives relations between Gibbs measure statistics on the \textit{same} closed surface, with different potentials.  
  
  Let us illustrate again with the $\C \cup \{\infty\}$ case. Let $\psi(z) = (a z + b)/(c z + d)$, $a,b,c,d \in \C$, $ad-bc = 1$ be a global conformal transformation of $\C \cup \{\infty\}$. Then LFT correlation functions transform as follows:
  \begin{equation}
  \left<  \prod_{i=1}^m \vertex_{a_i}(\psi(z_i)) \right>_b  =  \abs{\psi'(z_i)}^{-2\Delta_{a_i}}  \left< \prod_{i=1}^m \vertex_{a_i}(z_i)    \right>_b \,.
  \end{equation}
  Using \eqref{eq:generalSM}, the above yields the following result. Let $a_1, \dots, a_\ell < Q/2$ such that $a_1 + \dots  + a_\ell = Q$ be the charges. Let $U(z) = \sum_{j=1}^{\ell} 4 a_j \ln\abs{z - w_j}$ and $p_\beta(z) \propto e^{-\beta (\phi(z) + U(z))}$ be the  associated Gibbs measure; let also $V(z) = \sum_{j=1}^{\ell} 4 a_j \ln\abs{z - \psi(w_j)}$ and $\tilde{p}_\beta(z) \propto e^{-\beta (\phi(z) + V(z))}$. We used the convention $\ln\abs{z-\infty} = 0$ so that some $w_j$ and/or $\psi(w_j)$  can be $\infty$. Then we have 
  \begin{equation}
  \overline{\tilde{p}_\beta(\psi(z_1)) \dots \tilde{p}_\beta(\psi(z_n))} = \overline{p_\beta(z_1) \dots p_\beta(z_n)}
   \abs{\psi'(z_1)}^{-2} \dots \abs{\psi'(z_n)}^{-2} \label{eq:conformal}
  \end{equation}
 Although we do not know yet the proportion constant in \eqref{eq:generalSM}, we can check that the global factor of \eqref{eq:conformal} is right, by integrating both sides and using change of variable. An application will be given later \eqref{eq:log} to determine the asymptotic decay of $p_\beta(z)$ as $z \to \infty$.
    
\subsection{Conformal bootstrap of LFT and discrete terms}
 \subsubsection{Conformal bootstrap in general}
The conformal bootstrap approach aims to directly compute the 4-point correlation function $\left< \mathcal{V}_{a_1}(0)\mathcal{V}_{a_2}(1)\mathcal{V}_{a_3}(z)\mathcal{V}_{a_3}(\infty)\right>$ on the basis of general assumptions and consistency conditions. More details and references can be found in the review article \cite{ribault2014conformal} and in \cite{ribault2015liouville}. 

The symmetry group of any 2D conformal field theory is generated by two copies of Virasoro algebra with the same central charge $c$. The representations of these algebras, called Verma modules, are constructed by the primary field $\Phi_{\Delta, \bar{\Delta}}$ and its descendants, which are obtained by acting on the primary field with the symmetry generators. $\Phi_{\Delta, \bar{\Delta}}$ is characterized by the dimensions $\Delta$ and  $\bar{\Delta}$ (which is \textit{not} the complex conjugate $\Delta^*$) that give its scaling dimension $\Delta+\bar{\Delta}$ and its spin $\Delta-\bar{\Delta}$. The set of representations that form the space of fields of a CFT is called the spectrum $\mathcal{A}$. The most general form of a four-point correlation consistent with conformal symmetry takes the form:
\begin{align}
 \left< \Phi_{\Delta_1,\bar{\Delta}_1}(0)\Phi_{\Delta_4,\bar{\Delta}_4}(z)\Phi_{\Delta_2,\bar{\Delta}_2}(1)\Phi_{\Delta_3,\bar{\Delta}_3}(\infty)\right> &= \sum_{(\Delta,\bar{\Delta})\in\mathcal{A}} C(\Delta_1,\Delta_4,\Delta)C(
 \Delta,\Delta_2,\Delta_3)\mathcal{F}_{\Delta}(\{\Delta_i\},z) \mathcal{F}_{\bar{\Delta}}(\{\bar{\Delta}_i\},z) \, \label{genexpr}
\end{align} 
 The functions $ C(\Delta_1,\Delta_2,\Delta_3)$ and $\mathcal{F}_{\Delta}(\{\Delta_i\};z)$ are respectively the structure constants of the theory and the conformal block. The latter, $\mathcal{F}_{\Delta}(\{\Delta_i\},z)$, is a function of $\Delta, \Delta_i, z$ and $c$, and is \textit{universal for all CFTs}. In contrast, the spectrum $\mathcal{A}$ and the structure constants depend on the particular CFT in question and allows for the computation of its correlation functions by (\ref{genexpr}). 

\subsubsection{LFT by conformal bootstrap}
The spectrum $\mathcal{A}_L$ and the structure constants $C^{\text{DOZZ}}$ (DOZZ for Dorn-Otto and Zamolodchikov-Zamolodchikov, who found them independently \cite{dorn1994two,zamolodchikov1996conformal}) of the LFT can be derived as the only solution that satisfy four main assumptions: 

\begin{itemize}
\item \underline{Assumption 1:}  The spectrum is  {\it diagonal}, \textit{i.e.}, the primary fields are all spinless, $\Delta = \bar{\Delta}$. 

The primary fields of the LFT are the vertex operators $\mathcal{V}_{a}$, whose parameter $a$ is called the \textit{charge}, and whose dimension is $(\Delta_a,\Delta_a)$, where 
\begin{equation}
\Delta_a = a(Q - a) \,,
\end{equation}  
see \eqref{eq:ope} of the main text. Here $Q = b + 1 / b$ (see \eqref{eq:defQ} of the main text) is related to the central charge by $c = 1 + 6 Q^2$. LFT can be constructed for any value $c \in \C$ \cite{ribault2015liouville}, but here we will restrict to $c \geq 25$, corresponding to $Q \geq 2$ and $b \in (0,+\infty)$. 

\item \underline{Assumptions 2:} The spectrum contains {\it continuously} many Verma modules of the Virasoro algebra, each of them having multiplicity one. 

The sum in (\ref{genexpr}) is therefore meant to be an integral. When parametrized by the charge $a$, the measure of the integration is linear, $\sum_{\Delta_a \in \mathcal{A}_L} \to \int_{\mathcal{A}_L} \dif a $.

\item \underline{Assumption 3:} The theory is unitary. 

In terms of the charge $a$, the \textit{unitary bound} $\Delta \in (0, +\infty)$ becomes $a \in (0, Q) \cup \left(\frac{Q}{2} + \im \mathbb{R}\right)$.

\item \underline{Assumption 4:} The LFT correlation functions are single-valued functions of $\{z_i\}$, and depend smoothly on the central charge $c$ and the conformal dimensions of the fields.

The Assumption 4 requires the contour integral  $\int_{\mathcal{A}_L}\dif a$ to have no end-points. This fixes the LFT spectrum $\mathcal{A}_L$ to be: 
\begin{equation}
\label{LFTspect}
\mathcal{A}_L = \frac{Q}{2} + \im \R =  \left\lbrace \frac{Q}{2} + \im P: P \in \R \right\rbrace
\end{equation}
\end{itemize}  

These assumptions, combined with the conditions of crossing symmetry, are sufficient to solve completely the theory. 
In particular one has for the structure constants:
\begin{equation}
\label{DOZZ}
C^{\text{DOZZ}}(a_1,a_2,a_3)= \frac{\left[b^{\frac{2}{b}-2b} \mu\right]^{Q-a_1-a_2-a_3}\prod_{i=1}^{3}\Upsilon_b(2 a_i)}{\Upsilon_b(\sum_{i=1}^3 a_i-Q)\Upsilon_b( a_1+a_2-a_3)\Upsilon_b(a_1-a_2+a_3)\Upsilon_b(-a_1+a_2+a_3)},
\end{equation}
 The function $\Upsilon_b(x)$ can be defined by an integral formula, which is valid for $0<\Re x<\Re Q$, 
\begin{align}
 \log\Upsilon_b(x) = \int_0^\infty \frac{dt}{t} \left[\left(\tfrac{Q}{2}-x\right)^2 e^{-2t} -\frac{\sinh^2\left(\frac{Q}{2}-x\right)t}{\sinh bt\sinh\frac{t}{b}}\right]\,.
\label{lup}
\end{align}
and then continue to $x \in \C$ using the recursion relations
\begin{equation}
\frac{\Upsilon_b(x+b)}{\Upsilon_b(x)} = b^{1 - 2 b x} \gamma(bx) \,,\, 
\frac{\Upsilon_b(x+b^{-1})}{\Upsilon_b(x)} = b^{\frac{2x}b - 1} \gamma(x/b) \,,\, \gamma(x) \defeq \frac{\Gamma(x)}{\Gamma(1-x)} \,.
\label{eq:upprecursion}
\end{equation}
There is also the following product formula valid for all $x\in\mathbb{C}$:
\begin{align}
 \Upsilon_b(x) = \lambda_b^{(\frac{Q}{2}-x)^2}\prod_{m,n=0}^\infty f\left(\frac{\frac{Q}{2}-x}{\frac{Q}{2}+mb+nb^{-1}}\right) \quad \text{with} \quad f(x)=(1-x^2)e^{x^2}\,,\,\Upsilon_b(Q/2) = 1 \,.
\label{upp}
\end{align}
where $\lambda_b$ is a numerical constant. $\Upsilon_b$ is related to the $\beta$-Barnes function $G_\beta(x)$ by (\cite{fateev2000boundary}, eq. 3.16)
\begin{equation} \Upsilon_b(x) = G_b(x) G_{b}(Q-x) \,. \end{equation} 
See also \cite{fyodorov2016interval}, Appendix G (Sect. 13.2) for other normalizations.

$\Upsilon_b(x)$ is analytic on $\mathbb{C}$, with infinitely many simple zeros:
\begin{align}
 \Upsilon_b(x) = 0 \quad \Leftrightarrow \quad x \in &\left\{ -  b m - b^{-1} n : m,n= 0,1,2 \dots  \right\} \bigcup 
 \left\{ Q +  b m +  b^{-1} n : m,n= 0,1,2 \dots  \right\}  \nonumber \\
   = & (-L) \bigcup \, (Q + L) \,,\, L \defeq 
  \left\{b m +  b^{-1} n : m,n= 0,1,2 \dots \right\}  \,. \label{eq:Uppzero}
\end{align}
Observe they are organized into two \textit{lattices} (defined as some affine transform of $L$): one ranging from $0$ to $-\infty$, the other from $Q$ to $+\infty$, and the two related by the symmetry $a \mapsto Q - a$.
  \vspace{.2cm}

Equations \eqref{genexpr}, (\ref{LFTspect}) and (\ref{DOZZ}) imply the following expression of LFT $4$-point functions of primary operators that belong to the LFT spectrum:
 \begin{equation}
 \label{intcont}
\left< \mathcal{V}_{a_1}(0)\mathcal{V}_{a_4}(z)\mathcal{V}_{a_2}(1)\mathcal{V}_{a_3}(\infty)\right> = \int_{\im \mathbb{R}+ \frac{Q}{2}} C^{\text{DOZZ}}(a_1,a_4,a)C^{\text{DOZZ}}(Q-a,a_2,a_3)|\mathcal{F}_{\Delta_{a}}(\{a_i\},z)|^2\dif a  \,, 
\end{equation}
where $\Re(a_1), \dots, \Re(a_4) = Q / 2$. The second structure constant has $Q - a$ instead of $a$ for a reason related to the LFT \textit{reflection relation}, see \cite{ribault2014conformal}, Sect. 3.1 for detailed explanation.

For our application, $a_i \in (0, Q / 2)$ (see \eqref{eq:mainhighT}), so eq. \eqref{intcont} needs to be extended. The proper way to do this, so as to preserve crossing symmetry and the Assumptions, is known \cite{belavin2006integrals,aleshkin2016construction}, as we  recall below. 

\subsubsection{Poles structure and discrete terms}
As a function of $a \in \C$, the integrand of \eqref{intcont} has poles. They come from two origins:

The conformal block $\mathcal{F}_{\Delta_a}(\{a_i\},z)$, as a function of $a$, has poles at
$ - \frac12L$ (see \eqref{eq:Uppzero} for the definition of $L$), corresponding to degenerate fields flowing in the internal channel. They will not be important in our discussion.

In addition, the 2 structure constants $C^{\text{DOZZ}}(a_1, a_4, a)$ and $C^{\text{DOZZ}}(Q - a, a_2, a_3)$ in \eqref{intcont} have poles, coming from the zeros of the $\Upsilon$'s in the denominator of the DOZZ formula \eqref{DOZZ}.  As we have seen in \eqref{eq:Uppzero}, the $\Upsilon$ function has 2 lattices of zeros; the DOZZ formula contains 4 $\Upsilon$'s. In total, there are 16 lattices of poles, 8 for each structure constant. Let us focus on those of $C^{\text{DOZZ}}(a, a_1, a_4)$. The 8 lattices are as follows.
\begin{eqnarray}
&\begin{tabular}{|c|c|c|c|c|c|c|c|}
\hline
1 & 2 & 3 & 4 & 5 & 6 & 7 & 8 \\ \hline
 $Q - a_s - L$ &   $a_s  +  L$  &  $2Q - a_s - L$ & $-Q + a_s - L$ & 
 $a_d - L$ & $-a_d + Q + L$  & $-a_d - L$ & $a_d + Q + L$ \\ \hline
\end{tabular} \label{eq:tablepoles} \\
& \text{where }a_s \defeq a_1 + a_4 \,,\, a_d \defeq a_1 - a_4 \,,\,  L = \left\{b m +  b^{-1} n : m,n= 0,1,2 \dots \right\} \text{, see eq. \eqref{eq:Uppzero}.} \label{eq:asumdiff}
\end{eqnarray}
Note that the lattices $1$ and $2$ are related by $a \mapsto Q - a$ symmetry, so do $3$ and $4$, \textit{etc}. The poles coming from $C^{\text{DOZZ}}(Q-a, a_2, a_3)$ are obtained by replacing $1,4 \to 2,3$ in the above equation. 
\begin{figure}
\begin{center}
 \begin{tikzpicture}[scale = 0.5,  every node/.style={scale=1.1}]
\draw[red, very thick,postaction={decorate,decoration={markings,
    mark=at position .65 with {\arrow[scale=1]{stealth}}}}] (1, -6) -- (1, 6);
\draw[thick,->] (-6,0)--(6,0)node[right]{$\Re(a)$};;
\draw[thick,->] (0,-6)--(0,6) node[above]{$\Im(a)$};
\draw[thick] (2,-6)--(2,6);
\draw (-0.4,0) node[below]{$0$};
\draw (2+0.4,0) node[below]{$Q$};
\draw (1+0.4,0) node[below]{$\frac{Q}{2}$};
\draw (-7,3.5) node[right]{$a_1+a_4$};
\draw (-7,1.5) node[right]{$a_4-a_1$};
\draw (-7,-0.5) node[right]{$a_1-a_4$};
\draw (-7,-2.5) node[right]{$-a_1-a_4$};
\foreach \y in {3,-1,1,-3}
\foreach \x in {0,0.47619, 0.952381, 1.42857, 2.1, 2.57619, 3.05238}
\node[draw,circle,inner sep=1pt,fill] at (2+\x,\y) {};
\foreach \y in {3,-1,1,-3}
\draw[dashed] (5.07,\y)--(6,\y);
\foreach \y in {3,-1,1,-3}
\foreach \x in {0,0.47619, 0.952381, 1.42857, 2.1, 2.57619, 3.05238}
\node[draw,circle,inner sep=1pt,fill] at (-\x,\y) {};
\foreach \y in {3,-1,1,-3}
\draw[dashed] (-3.1,\y)--(-6,\y);
\foreach \y in {3,-1,1,-3}
\foreach \x in {0,0.47619, 0.952381, 1.42857, 2.1, 2.57619, 3.05238}
\node[draw,circle,inner sep=1pt,fill] at (2+\x,\y) {};
\foreach \y in {3,-1,1,-3}
\draw[dashed] (5.07,\y)--(6,\y);
\foreach \y in {3,-1,1,-3}
\foreach \x in {0,0.47619, 0.952381, 1.42857, 2.1, 2.57619, 3.05238}
\node[draw,circle,inner sep=1pt,fill] at (-\x,\y) {};
\foreach \y in {3,-1,1,-3}
\draw[dashed] (-3.1,\y)--(-6,\y);
\draw (-7,5) node[right]{$a_2+a_3$};
\draw (-7,2.5) node[right]{$a_2-a_3$};
\draw (-7,-1.5) node[right]{$a_3-a_2$};
\draw (-7,-4) node[right]{$-a_2-a_3$};
\foreach \y in {4.5,-2,2,-4.5}
\foreach \x in {0,0.47619, 0.952381, 1.42857, 2.1, 2.57619, 3.05238}
\node[draw,circle,inner sep=1pt,fill] at (2+\x,\y) {};
\foreach \y in {4.5,-2,2,-4.5}
\draw[dashed] (5.07,\y)--(6,\y);
\foreach \y in {4.5,-2,2,-4.5}
\foreach \x in {0,0.47619, 0.952381, 1.42857, 2.1, 2.57619, 3.05238}
\node[draw,circle,inner sep=1pt,fill] at (-\x,\y) {};
\foreach \y in {4.5,-2,2,-4.5}
\draw[dashed] (-3.1,\y)--(-6,\y);
\foreach \x in {0,0.47619, 0.952381, 1.42857, 2.1, 2.57619, 3.05238}
\node[draw,circle,inner sep=1pt] at (-\x,0) {};
\draw[dashed, thick] (-3.1,0)--(-6,0);
 \end{tikzpicture}
  \begin{tikzpicture}[scale = 0.5,  every node/.style={scale=1.1}]
 \draw[red, very thick,postaction={decorate,decoration={markings,
     mark=at position .65 with {\arrow[scale=1]{stealth}}}}] (1, -6) -- (1, 6);
 \draw[thick,->] (-6,0)--(6,0)node[right]{$\Re(a)$};;
 \draw[thick,->] (0,-6)--(0,6) node[above]{$\Im(a)$};
 \draw[thick] (2,-6)--(2,6);
 \draw (-0.4,0) node[below]{$0$};
 \draw (2+0.4,0) node[below]{$Q$};
 \draw (1+0.4,0) node[below]{$\frac{Q}{2}$};
 \draw (-7,3.5) node[right]{$a_1+a_4$};
 \draw (-7,1.5) node[right]{$a_4-a_1$};
 \draw (-7,-0.5) node[right]{$a_1-a_4$};
 \draw (-7,-2.5) node[right]{$-a_4-a_1$};
 \foreach \y in {3}
 \foreach \x in {0,0.47619, 0.952381, 1.42857, 2.1, 2.57619, 3.05238}
 \node[draw,circle,inner sep=1pt,fill] at (0.3+\x,\y) {};
 \draw[red] (0.3,3) circle (0.18cm);
 \draw[red] (0.3+0.47619,3) circle (0.18cm);
 \foreach \y in {-1,1}
 \foreach \x in {0,0.47619, 0.952381, 1.42857, 2.1, 2.57619, 3.05238}
 \node[draw,circle,inner sep=1pt,fill] at (2+\x,\y) {};
 \foreach \y in {-3}
 \foreach \x in {0,0.47619, 0.952381, 1.42857, 2.1, 2.57619, 3.05238}
 \node[draw,circle,inner sep=1pt,fill] at (3.7+\x,\y) {};
 \foreach \y in {3,-1,1,-3}
 \draw[dashed] (5.07,\y)--(6,\y);
 \foreach \y in {3}
 \foreach \x in {0,0.47619, 0.952381, 1.42857, 2.1, 2.57619, 3.05238}
 \node[draw,circle,inner sep=1pt,fill] at (-1.7-\x,\y) {};
 \foreach \y in {-1,1}
 \foreach \x in {0,0.47619, 0.952381, 1.42857, 2.1, 2.57619, 3.05238}
 \node[draw,circle,inner sep=1pt,fill] at (-\x,\y) {};
 \foreach \y in {-3}
 \foreach \x in {0,0.47619, 0.952381, 1.42857, 2.1, 2.57619, 3.05238}
 \node[draw,circle,inner sep=1pt,fill] at (1.7-\x,\y) {};
 \draw[red] (1.7,-3) circle (0.18cm);
 \draw[red] (1.7-0.47619,-3) circle (0.18cm);
 \foreach \y in {3,-1,1,-3}
 \draw[dashed] (-3.1,\y)--(-6,\y);
 \draw (-7,3.5) node[right]{$a_1+a_4$};
 \draw (-7,1.5) node[right]{$a_4-a_1$};
 \draw (-7,-0.5) node[right]{$a_1-a_4$};
 \draw (-7,-2.5) node[right]{$-a_4-a_1$};
 \foreach \y in {3,-1,1,-3}
 \draw[dashed] (-3.1,\y)--(-6,\y);
 
 \draw (-7,5) node[right]{$a_2+a_3$};
 \draw (-7,2.5) node[right]{$a_3-a_2$};
 \draw (-7,-1.5) node[right]{$a_2-a_3$};
 \draw (-7,-4) node[right]{$-a_2-a_3$};
 \foreach \y in {4.5,-2,2,-4.5}
 \foreach \x in {0,0.47619, 0.952381, 1.42857, 2.1, 2.57619, 3.05238}
 \node[draw,circle,inner sep=1pt,fill] at (2+\x,\y) {};
 \foreach \y in {4.5,-2,2,-4.5}
 \draw[dashed] (5.07,\y)--(6,\y);
 \foreach \y in {4.5,-2,2,-4.5}
 \foreach \x in {0,0.47619, 0.952381, 1.42857, 2.1, 2.57619, 3.05238}
 \node[draw,circle,inner sep=1pt,fill] at (-\x,\y) {};
 \foreach \y in {4.5,-2,2,-4.5}
 \draw[dashed] (-3.1,\y)--(-6,\y);
 \foreach \x in {0,0.47619, 0.952381, 1.42857, 2.1, 2.57619, 3.05238}
 \node[draw,circle,inner sep=1pt] at (-\x,0) {};
 \draw[dashed, thick] (-3.1,0)--(-6,0);
  \end{tikzpicture}
 \caption{\textit{Left}: The poles of the integrand in (\ref{intcont}) in the complex plane of $a$ when $\Re(a_i)=\frac{Q}{2}$, $i=1,..,4$. The filled and empty dots indicate respectively the position of the poles of the structure constants \eqref{eq:tablepoles} and of the conformal blocks ($-L/2$). The red line is the integration contour $\mathcal{A}_L = Q / 2 + \im \R$. \textit{Right}:
 The same plot, with a parameter set such that $\Re(a_2) = \Re(a_3) = Q / 2$ but $\Re(a_1) = \Re(a_4) < Q / 2$, and $\Re(a_1 + a_4) < Q/2$. The poles that have crossed the contour are marked by red circles.} 
 \label{poles1}
 \end{center}
\end{figure}
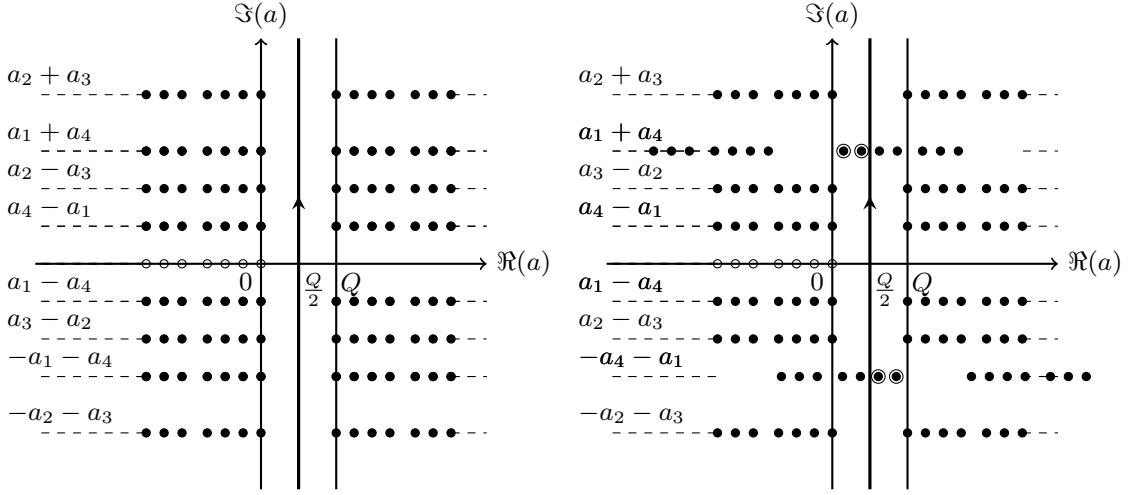

\begin{figure}
  \begin{center}
    (i) \includegraphics[scale=0.45,valign=t]{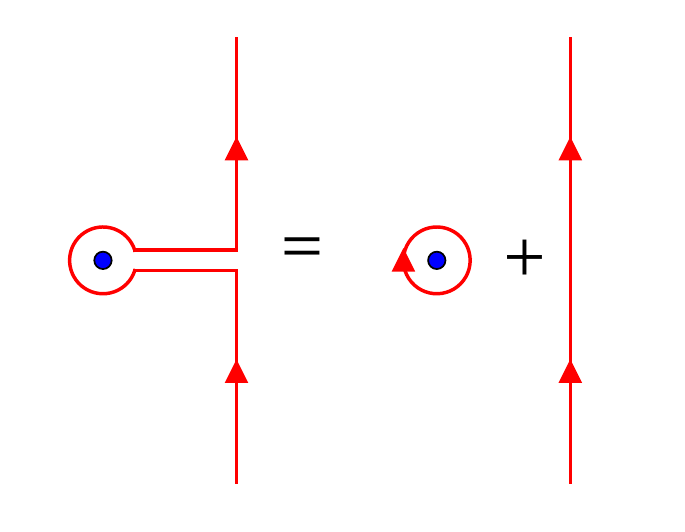}
    (a)\includegraphics[scale=.4,valign=t]{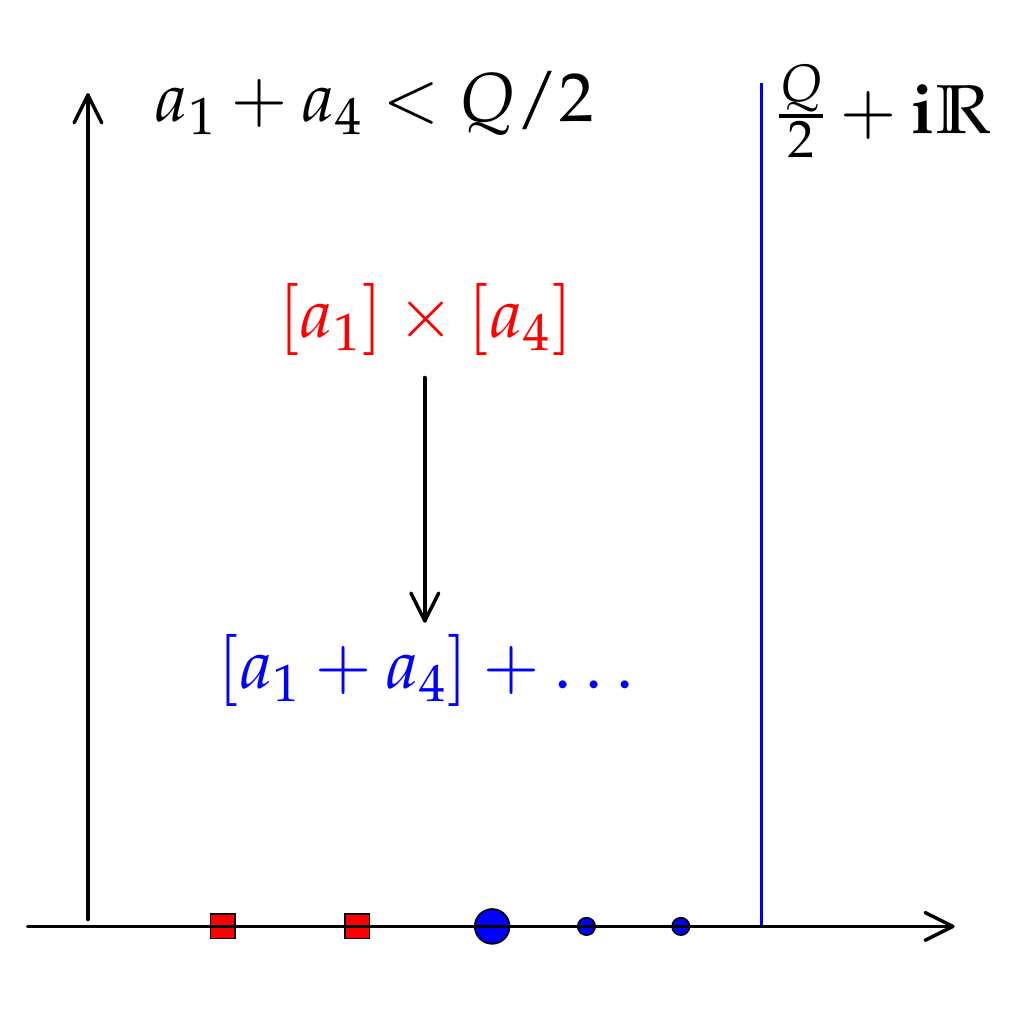}
    (b)\includegraphics[scale=.4,valign=t]{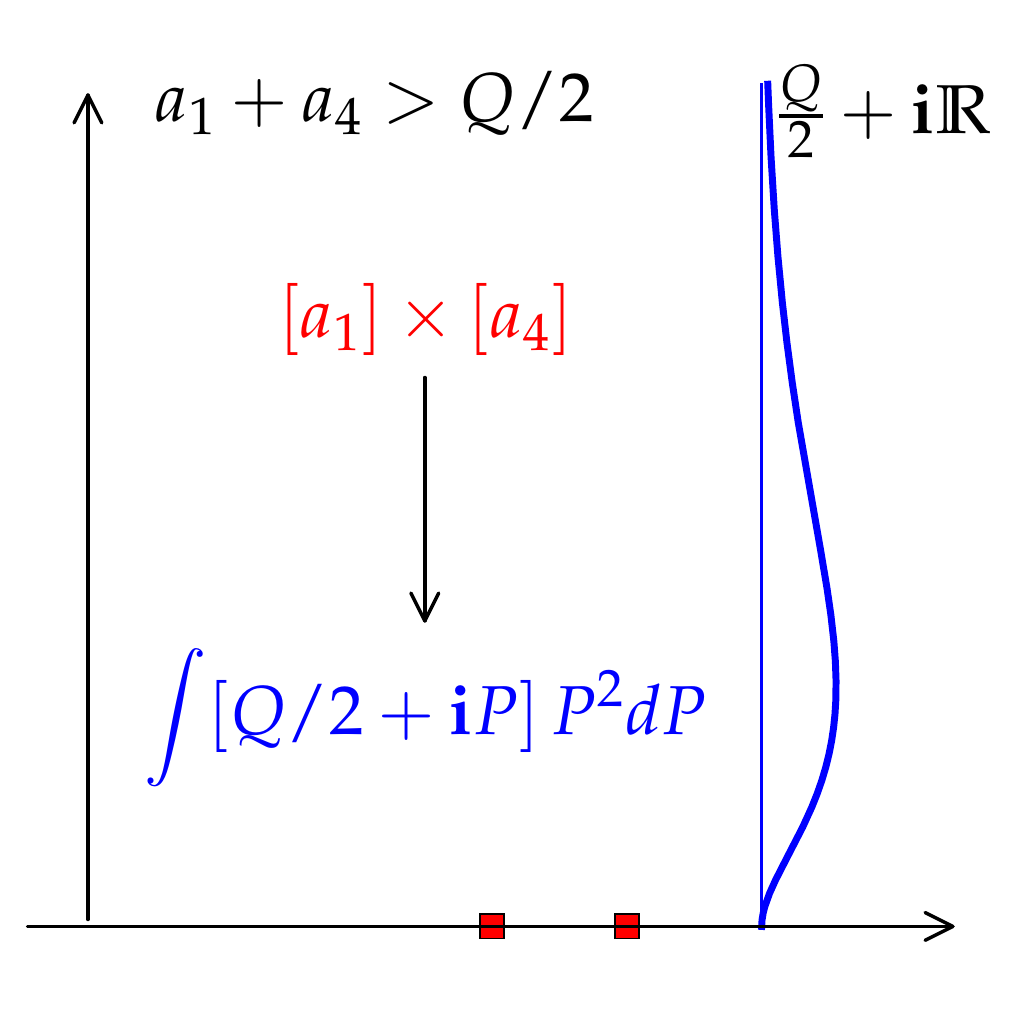}
    (c)\includegraphics[scale=.4,,valign=t]{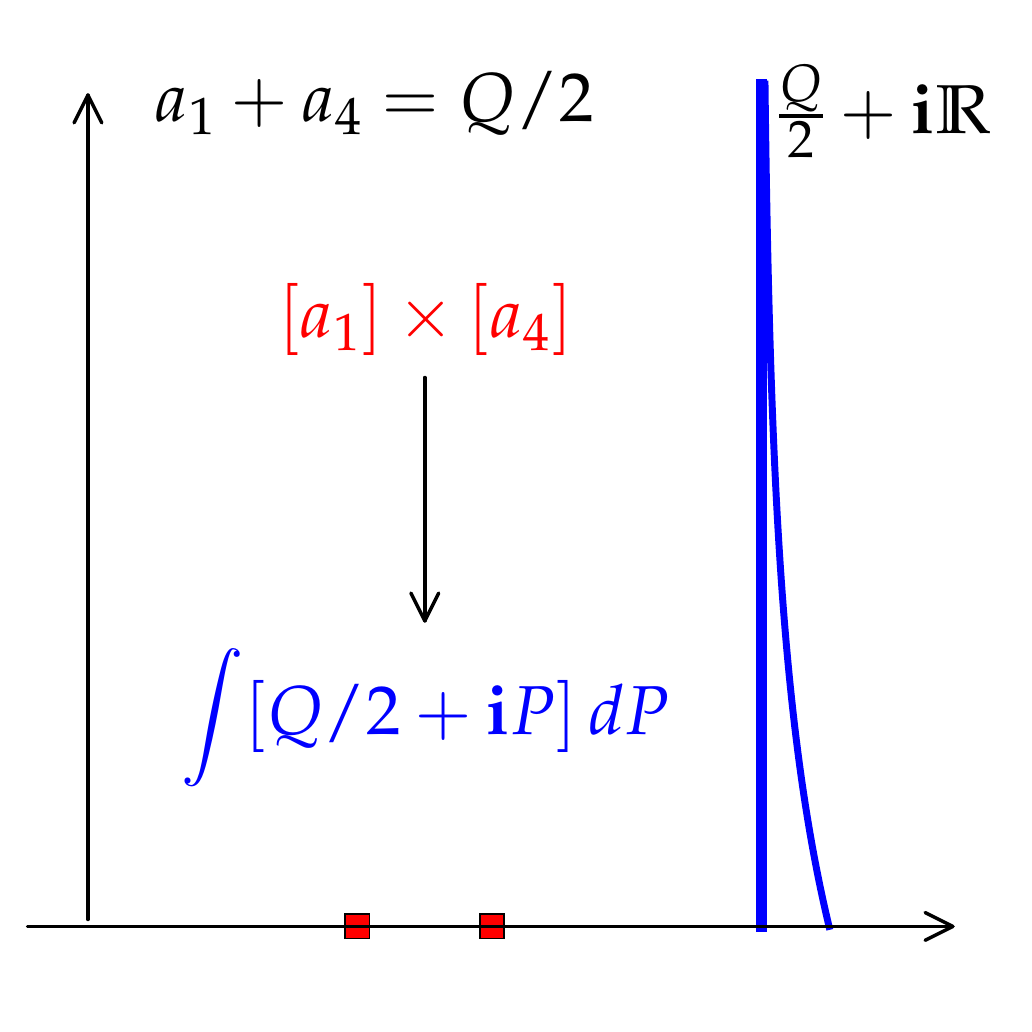}
  \caption{(i): Deformation of the contour line when a pole in $P_-$ (black dot) have crossed the line $Q / 2 + \im P$ (red vertical line) from its right. The orientation of the surrounding circle implies that the residue contribution is $\mathbf{-} 2\pi \im \mathrm{Res} [\dots]$. (a-c) Illustrations of the three cases of OPE. In each case the (upper-half) complex plane of charge values are drawn. Red squares indicate the charges $a_1$ and $a_4$. The charge(s) dominating the OPE is written in blue. The Blue dots indicate the positions of poles included in \eqref{intpdis}; the largest one dominates the OPE. The blue straight line is the LFT spectrum. The blue thick curves are cartoons of the value of the integrand in \eqref{intcont}.}  \label{contour}
  \end{center}
   \end{figure}

If the $a_i \in \mathcal{A}_L$, \textit{i.e.} $\Re(a_i) = \frac{Q}{2}$, the real part of the poles belongs to the intervals $(-\infty, 0] \cup [Q, +\infty)$, which do not intersect the integration contour $Q/2 + \im \R$ (\ref{intcont}), as shown in Figure \ref{poles1} (Left). Now, when $\Re(a_s) = \Re(a_1+a_4)$ starts to decrease from $Q$ into the interval $(0,Q)$, the lattices of poles start to move on the plane. A case-by-case check using \eqref{eq:tablepoles} shows that only the lattices $1$ and $2$ may cross the line $Q/2 + \im \R$. When $\Re(a_s)$ decreases to $Q / 2$, lattice $1$ crosses the line from its left, and so does lattice $2$ from it right. As $\Re(a_s)$ further decreases, several poles from those lattices will have crossed the line. These poles are: 
\begin{align}
 P_+ \defeq \left\{ x \in Q - (a_1 + a_4) - L : \Re(x) \in  (Q/2, Q) \right\} \,,\, P_- \defeq \left\{ x \in a_1 + a_4 + L : \Re(x) \in  (0, Q / 2) \right\} 
 \,. \label{eq:includepoles}
\end{align}
 Figure \ref{poles1} (Right) illustrates such a situation. In order to extend analytically the integral (\ref{intcont}), the integration contour has to be deformed so as to avoid the poles from crossing it. By Cauchy formula (applied in the fashion illustrated in Figure \ref{contour} (i)), this amounts to adding $ \pm 2\pi \im$ times the residues of the integrand of \eqref{intcont} at points in $P_\pm$, respectively. They are the so called \textit{discrete terms}. 
 It is not hard to show, using the $a \mapsto Q - a$ symmetry of $\Upsilon$,  that the contribution of $P_+$ equals that of $P_-$. Finally, the poles from $a_{2,3}$ can be similarly treated.
 The resulting expression for LFT  $4$-point function for values of $\Re(a_i)\in (0,Q/2) $ is as follows:
  \begin{eqnarray}
  \label{intpdis}
  \left< \mathcal{V}_{a_1}(0)\mathcal{V}_{a_2}(1)\mathcal{V}_{a_4}(z)\mathcal{V}_{a_3}(\infty)\right> = \int_{\frac{Q}{2} + \im \mathbb{R}} C^{\text{DOZZ}}(a_1,a_4,a)C^{\text{DOZZ}}(Q-a,a_2,a_3)|\mathcal{F}_{\Delta_{a}}(\{a_i\},z)|^2\dif a +\nonumber \\
  -   2 \sum_{p \in P_-} 2\pi \im \;\text{Res}_{a \to p} \left[C^{\text{DOZZ}}(a,a_1,a_4) C^{\text{DOZZ}}(Q - a,a_2,a_3)\right]  \abs{\mathcal{F}_{\Delta_{a}}(\{a_i\},z)}^2 + [(1,4) \leftrightarrow (2,3)] \,.
  \end{eqnarray}
 Eq. \eqref{intpdis} is the formula that is implemented for the numerical tests (plotted as ``LFT'' curves in Figure \ref{fig:tests} in the main text). The code-base \cite{codebootstrap} implements already the continuous integral part (or \eqref{intcont}). For the discrete terms part, note that the residues of the $\Upsilon$ function (thus those of the DOZZ formula) can be exactly obtained by the recursion relations \eqref{eq:upprecursion}. We checked to high numerical precision that crossing symmetry is satisfied by \eqref{intpdis} (but \textit{not} by \eqref{intcont}) in the parameter range of our interest (\textit{i.e.}, $a_i \in (0,Q/2)$).

\subsubsection{Asymptotic behaviour (OPE)}
We consider the asymptotic behaviour of eq. \eqref{intpdis} as $z \to 0$, assuming $a_i \in (0, Q / 2)$. For this, note that the $z$-dependence of the $4$-point function \eqref{intpdis} come from the conformal blocks. The $z\to 0$ series expansion of the latter is well-known:
\begin{equation}
\label{cbexp}
\mathcal{F}_{\Delta_a}(\{\Delta_{a_i}\},z) = z^{-\Delta_{a_1}-\Delta_{a_4}+\Delta_a} \left(1+\frac{(\Delta_{a_4}-\Delta_{a_1}+ \Delta)(\Delta_{a_2}-\Delta_{a_3}+ \Delta)}{2\Delta} z + O(z^2)\right) \,.
\end{equation}
Therefore, to compute the dominant asymptotic behaviour of \eqref{intpdis} as $z\to0$, we need to consider the internal charges $a \in P_+ \cup (Q/2 + \im \R)$ involved, and find the \textit{smallest} scaling dimension $\Delta_a = a(Q - a)$. We have to distinguish three cases, illustrated in Figure \ref{contour}, (a)-(c):
\begin{itemize}
\item[(a)] $a_1+a_4< \frac{Q}{2}$ (pole crossing).

 The smallest scaling dimension is given by the discrete term $p = a_1 + a_4 \in P_-$. So \eqref{cbexp} implies
\begin{equation}
\label{disbehav}
 \left< \mathcal{V}_{a_1}(0)\mathcal{V}_{a_4}(z)\mathcal{V}_{a_2}(1)\mathcal{V}_{a_3}(\infty)\right> 
\underset{z\to 0}{\sim} |z|^{-2 \delta_0} \,, \,  
\delta_0 = \Delta_{a_1} + \Delta_{a_4} - \Delta_{a_1 + a_4} = 2 a_1 a_4 \,.
\end{equation}
Note that the pole $p' = Q - a_1 - a_4 \in P_+$ ( see \eqref{eq:includepoles}) give the same contribution.

\item[(b)] $a_1+a_4> \frac{Q}{2}$ (no pole crossing).

There are no discrete terms, and the smallest dimension is given by $a= Q / 2$ in the continuous integral. One can check using \eqref{DOZZ} that both $C^{DOZZ}(a, a_1, a_4)$ and $C^{DOZZ}(Q-a, a_2, a_3)$ have a simple zero at $a = Q / 2$, so \eqref{cbexp} and \eqref{intpdis} imply
\begin{align}
&\label{intbehav}
 \left< \mathcal{V}_{a_1}(0)\mathcal{V}_{a_4}(z)\mathcal{V}_{a_2}(1)\mathcal{V}_{a_3}(\infty)\right> 
\underset{z\to 0}{\sim}  \int_{\mathbb{R}} |z|^{-2\delta_1 - 2P^2} P^2 \, \dif P  \sim |z|^{-2\delta_1} \ln^{-\frac32}|1/z| \,, \delta_1 = \Delta_{a_1} + \Delta_{a_4} - \Delta_{Q/2}  \,.
\end{align}

\item[(c)] $a_1+a_4 = \frac{Q}{2}$ (marginal case). 

This case is similar to the above one, except that $C^{DOZZ}(a, a_1, a_4)C^{DOZZ}(Q-a, a_2, a_3)$ does not vanish at $a = Q/2$, so we have 
\begin{equation}
\label{marginal}
 \left< \mathcal{V}_{a_1}(0)\mathcal{V}_{a_4}(z)\mathcal{V}_{a_2}(1)\mathcal{V}_{a_3}(\infty)\right> 
\underset{z\to 0}{\sim}  \int_{\R} |z|^{-2\delta_0 - 2P^2} \dif P  \sim |z|^{-2\delta_0} \ln^{-\frac12}|1/z|\,.
\end{equation}
This case can also be understood as the pair of dominant poles in case (a) merging at $Q / 2$ and compensating the double zero of case (b).
\end{itemize}
The results \eqref{disbehav}, \eqref{intbehav} and \eqref{marginal} can extend to general LFT $n$-point functions (except for correlation functions of $n\leq3$ points on sphere like surfaces),  leading to eq. \eqref{eq:ope} in the main text. 
Therefore, the termination point transition corresponds to the (first) crossing of poles of the DOZZ-structure constant through the LFT spectrum. The \textit{absence} discrete terms make the continuous integral dominate the OPE and lead to the log corrections. Their exponents $\frac{1}{2}$ and $\frac{3}{2}$ come from the vanishing order the DOZZ structure constants at $Q/2$. It should be stressed that no log-CFT is involved. 

We provide explicit formulae for the two applications in the main text:
\begin{enumerate}
\item $a_1, a_4 = a_1, b$ (note $b = \min(\beta, 1)$), corresponding to the divergence of the the Gibbs measure near a log-singularity of the potential:
\begin{align}
\overline{p_\beta(z)} \stackrel{z\to0}{\sim} \begin{cases}   
  \abs{z}^{-4 a_1 b}  & a_1+ b <  Q/2 \\
  \abs{z}^{2b^2 - 2} \ln^{-\frac{1}{2}}|1/z|  & a_1 + b = Q/2 \\
\abs{z}^{\frac{(Q - 2a_1)^2}{2}-2} \ln^{-\frac{3}{2}}|1/z|  & a_1 + b > Q/2 
 \end{cases} \label{eq:log}
\end{align}
We can apply the conformal covariance \eqref{eq:conformal} with $\psi(z) = 1 / z$ to study $p_\beta(z)$ at $z\to \infty$ in the $\C\cup\{\infty\}$ geometry. We stick to the minimal setting of \eqref{eq:mainhighT} (generalizations are obvious). Noting $a_3 = Q - a_1 - a_2$ , we have 
\begin{equation}
\overline{p_\beta(z)} \stackrel{\abs{z}\to+\infty}{\sim} \begin{cases}   
  \abs{z}^{4 a_3 b - 4}  & a_3 + b <  Q/2 \\
  \abs{z}^{-2 - 2b^2} \ln^{-\frac{1}{2}}|z|  & a_3 + b = Q/2 \\
\abs{z}^{\frac{-(Q - 2a_3)^2}{2}-2} \ln^{-\frac{3}{2}}|z|  & a_3 + b > Q/2 
 \end{cases} \label{eq:loginfty}
\end{equation}
\item $a_1, a_4 = b, b$, corresponding to the divergence of two point Gibbs measure correlation:
\begin{equation}
\overline{p_\beta(w) p_\beta(z + w)} \stackrel{z\to0}{\sim}
\begin{dcases}
\abs{z}^{-4 \beta^2} &  \beta < 3^{-\frac{1}{2}}  \\
\abs{z}^{-4/3} \ln^{-\frac{1}{2}}\abs{1/z} & \beta = 3^{-\frac{1}{2}} \\
\abs{z}^{-3 + \frac{\beta^2 + \beta^{-2}}{2}} \ln^{-\frac{3}{2}}\abs{1/z} &  \beta \in (3^{-\frac{1}{2}},1] \\
 c' T \abs{z}^{-2} \ln^{-\frac{3}{2}}\abs{1/z} + (1-T) \delta(z) & \beta > 1 
\end{dcases} \label{eq:pbpb}
\end{equation}
where in the last equation, $T = 1/\beta$ and $c'$ is a $T$-independent constant.
\end{enumerate}

\subsection{LogREM basics}
  \subsubsection{Replica symmetry breaking (RSB)}
  This section sketches the derivation of the result \eqref{eq:onestep} in the main text.
  To better illustrate the RSB technique, we consider a slightly more general case. Let $f(z)$ and $g(z)$ be two arbitrary functions defined for $z \in \C$, and suppose $\zeta_1, \zeta_2$ are the positions of two independent thermal particles at temperature $1/\beta$ in a same random potential; we wish to calculate the average over disorder $ \overline{f(\zeta_1) g(\zeta_2) }$. The standard replica trick calculates such observables by the following limit (we denote $V(z) = \phi(z) + U(z)$ the potential)
      \begin{align}
      &\overline{f(\zeta_1) g(\zeta_2)} = \lim_{n\to 0} \overline{f(\zeta_1)g(\zeta_2) Z^n} \,,\, \overline{f(\zeta_1)g(\zeta_2) Z^n} = \int \dif^2 \zeta_1 \dots  \dif^2 \zeta_n \,\overline{e^{-\beta V(\zeta_1)} \dots e^{-\beta V(\zeta_n)}} f(\zeta_1) g(\zeta_2)  \,. \label{eq:replicatrick}
      \end{align}
  That is, we integrate over $n$ \textit{replica positions} $\zeta_1, \dots, \zeta_n$. The crucial result of RSB is that the above integral runs over all possible positions $(z_1, \dots, z_n) \in \C^n$ \textit{only} when $\beta \leq 1$. When $\beta > 1$, the $n$ replicas are constrained to form $n /m$ tightly bound groups, each of size $m$, where $m = T =  1/\beta$, see \cite{fyodorov2010freezing}, or \cite{cao16order}, Section IV.A. So, \eqref{eq:replicatrick} should be written as a sum over all the replica-grouping configurations, and an integral over only $n/m$ group positions. Thanks to permutation symmetry, the sum amounts to some counting, and the combinatorics is worked out in for example in Appendix B of \cite{cao16order}. The result is the following:
      \begin{align}
      &\overline{f(\zeta_1)g(\zeta_2) Z^n} 
     = C_{n,m} \frac{m-1}{n-1}\left.\overline{f(\zeta_1) g(\zeta_1)Z^n}\right\vert_{\beta=1} + C_{n,m} \frac{n-m}{n-1}\left.\overline{f(\zeta_1) g(\zeta_2)Z^n}\right\vert_{\beta=1} \,.
      \end{align}
      Here, $C_{n,m}$ is a combinatorial factor (see eq. (58) of \cite{cao16order}) such that $C_{n,m}\to 1$ when $n \to 0$. In the last line, inside the averages $\left.\overline{[\dots]}\right\vert_\beta$ means that $\zeta_1$ and $\zeta_2$ are two independent thermal particles at the \textit{critical temperature $\beta=1$}. Now taking the $n\to0$ limit gives 
      \begin{equation}  \overline{f(\zeta_1)g(\zeta_2)} =  (1 - T) \left.\overline{f(\zeta_1) g(\zeta_1)}\right\vert_{\beta=1} +  T\left.\overline{f(\zeta_1)g(\zeta_2)}\right\vert_{\beta=1} \,,\, T = 1/\beta < 1 \,. \end{equation}
      Taking $f(z) = \delta(z-z_1)$ and $g(z) = \delta(z-z_2)$ gives eq. \eqref{eq:onestep} in the main text. 
 
 \subsubsection{Structure of deepest minima} 
 The deepest minima of a discrete 2D GFF \cite{biskup2016full} (and of a general logREM) are known to display a \textit{clustering} structure. Each cluster contains several deepest minima that are at $\epsilon$ distance from each other, and thus have the same position in the $\epsilon\to 0$ limit. At the level of first and second minimal values $V_{1,2}$ and positions $\xi_{1,2}$, this means two possibilities:
 \begin{enumerate}
 \item They come from a same cluster. Under this condition, the distribution of the common position $\xi_1 = \xi_2$ is given by $\overline{p_1(\xi_1)}$, \textit{i.e.}, the average Gibbs measure at the critical temperature.
 \item They come from two different clusters. Under this condition, the joint distribution of the positions $\xi_{1,2}$ is given by 
 $\overline{p_1(\xi_1) p_1 (\xi_2)}.$
 \end{enumerate}
Denoting $c_0$ the probability of event 1, we obtain \eqref{eq:min12} of the main text. Now to determine which of the two events happens, one needs to compare the two corresponding candidates for the second minimum. The first is in the same cluster as the minimum, let its value be $V_1 + v$. The distribution $P(v)$ of $v>0$  depends on the details of the 2D GFF at $\sim \epsilon$ scale (\textit{via} the \textit{decoration process}, see for example \cite{cao16order}, Appendix C and Section II.B.4). The second candidate is from another cluster; its value is $V_1 + \Delta$,  where $\Delta$ has the standard exponential distribution, and is independent of $v$. Now the true second minimum value is the smallest among the two candidates: $V_2 = V_1 + g, g = \min(v, \Delta)$: event 1 happens if and only if $v < \Delta$. So its probability is 
 \begin{equation} 
c_0 = \int_0^{+\infty} \dif v \int_v^{+\infty} \dif \Delta \, P(v) e^{-\Delta}
    =  \int_0^{+\infty} e^{-v} P(v) \dif v \,.
 \end{equation}
 It can be checked to be equal to $1- \overline{g}$, where the mean value of the gap is:
\begin{equation}
\overline{g} = \int_0^{+\infty} \dif v \int_0^{+\infty} \dif \Delta \, P(v) e^{-\Delta} \min(v, \Delta) = 1-c_0\,. 
\end{equation}

\subsubsection{From 2D to Cayley tree}
\begin{figure}
\includegraphics[scale=.6]{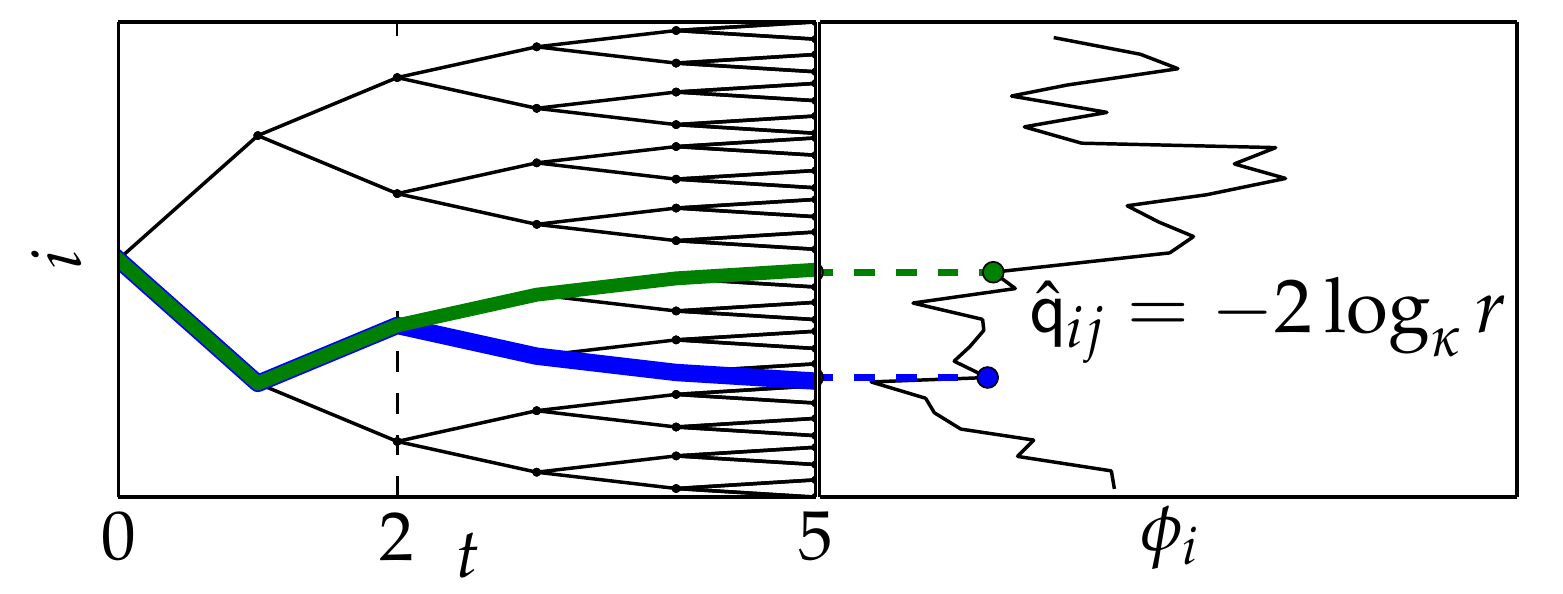}
\caption{A Cayley tree with $\kappa=2$ and $t=5$. Two directed polymers are drawn in bold; they have common length $\hat{\mathsf{q}}=2$. The energies of the DP's are plotted on the right. The common length--distance mapping is illustrated.}\label{fig:tree}
\end{figure}
Consider the logREM defined by a discrete 2D GFF $\phi(z)$ on a doubly periodic lattice (without potential $U = 0$). Since we will be interested only in short-distance behaviour, let's fix $R = 1$ and let $\epsilon \to 0$ so that the system size (lattice point number) is $ M = \epsilon^{-2}$. Let $K(z,w) = \overline{\phi(z)\phi(w)}^c$ denote the covariance of the 2D GFF (see \eqref{eq:gff} in the main text). Now let $z_1, z_2$ be the (random) position of two independent thermal particles (at temperature $\beta$) in one such random potential.

On the other hand, consider the Cayley tree model defined in the main text (see also Figure \ref{fig:tree}), also of size $M$, and denote its energy covariance by $C_{i,j} = \overline{\phi_i \phi_j}, i,j = 1, \dots, M$. Again, let $i_1, i_2$ be two independent thermal DP's on one disordered tree, also at temperature $\beta$. 

Observe that the two covariances have the same value range: $C_{ij}, K(z,w) \in [0, 2 \ln M = 2 t\ln \kappa]$. Now the conjectural relation is that, after averaging over all samples, the probability distributions of covariances $C_{i_1,i_2}$ and $K(z_1, z_2)$, have  the same behaviour in the ``scaling range'' \textit{i.e.},
\begin{equation}
\overline{\theta(C_{i_1, i_2} - y)}^{\text{DP}} = \overline{\theta(K(z_1,z_2) - y)}^{\text{2D}} \,,\,  1 \ll y \ll 2 \ln M \,, \label{eq:matching1}
\end{equation}
where $\theta$ is the Heaviside theta function. Now using \eqref{eq:gff} and \eqref{eq:covtree} in the main text we have (letting $\hat{\mathsf{q}} = y / (2\ln \kappa)$, and recalling $\hat{\mathsf{q}}_{ij} = C_{ij} / (2\ln \kappa)$)
\begin{equation}
\overline{\theta(\hat{\mathsf{q}}_{i_1i_2} - \hat{\mathsf{q}})}^{\text{DP}} = \overline{\theta(\kappa^{-\hat{\mathsf{q}}/2} - \abs{z_1 - z_2})}^{\text{2D}} 
= \int_{\abs{z}} \overline{p_{\beta}(w)p_\beta(w + z)\theta(\kappa^{-\hat{\mathsf{q}}/2} - \abs{z})}^{\text{2D}}  \,\dif^2 z \,,\, 
  0 \ll \hat{\mathsf{q}} \ll t
\end{equation}
 where we used the translation invariance property of the 2D GFF logREM model ($w$ is arbitrary). Observe that the matching of covariance induces the correspondence of the distances $r = \abs{z} = \kappa^{-\hat{q}/2}$ advocated in the main text. Taking $\dif / \dif \hat{\mathsf{q}}$ gives the transformation law of the pdf's of the main text. 
 
  \subsubsection{Correpondence between different logREMs}
  LogREMs can be defined on $d$-dimensional lattice as well as on hierarchical lattices. The notion of distance is clearly different from one to another, but can be transformed by comparing the covariance of the random potential. We refer to \cite{cao16order}, Sect. II.A for more background knowledge. 
  
  To do this conveniently, we will normalize the covariance so that the freezing temperature is always $\beta = 1$. For a general discrete logREM with potential values $V_1, \dots, V_M$, this can be done by requiring $\overline{V_{i}^2}^d = 2 \ln M + O(1)$ ($\overline{[\dots]}^d$ denotes the logREM average in $d$-dimension). For a logREM in $d$-dimension lattice (with lattice points $\mathbf{x}_i,i=1,\dots,M$), the covariance is normalized as
  \begin{equation}
  \overline{V_i V_j}^d = 2 d \ln \abs{R / (\mathbf{x}_i - \mathbf{x}_j)} \,, \, \epsilon \ll \abs{\mathbf{x}_i - \mathbf{x}_j} \ll R \,,
  \end{equation}
  where $\epsilon$ and $R$ are the lattice spacing and size, respectively, such that $M = (R/a)^d$ ($\overline{V_i}^d = 0$). The $d=2$ case is the 2D GFF model studied in the main text. A few $d=1$ cases are also widely studied in the literature thanks to their exact solvability. 
  
  For the moment, we cannot generalize the full mapping to LFT into dimensions other than $d=2$. Nevertheless, the asymptotic behaviours obtained by the Liouville OPE can be used to obtain analogous results for logREMs in any dimension. For this, we can use the same method as in the previous subsection, \textit{i.e.}, we identify the distance in $d$-dimension, $r_d$, to that in 2D, by $r_d^d = r_2^2$. 
  
  As an illustration, we apply this transformation law to eq. \eqref{eq:pbpb}. Taking into account the Jacobian, we have $P(r_2) \dif r_2^2 \propto P(r_d) \dif r_d^d$, where $P(r_2) = \overline{p_\beta(z) p_\beta(z + r_2)}$ in eq. \eqref{eq:pbpb} and $P(r_d) = \overline{p_\beta(0) p_\beta(\mathbf{x})}^{d}$. The result is:
 \begin{equation}
 \overline{p_\beta(0) p_\beta(\mathbf{x})}^d \stackrel{\abs{\mathbf{x}}\to0}{\sim}
 \begin{dcases}
 \abs{\mathbf{x}}^{-2d \beta^2} &  \beta < 3^{-\frac{1}{2}}  \\
 \abs{\mathbf{x}}^{-2d/3} \ln^{-\frac{1}{2}}\abs{1/\mathbf{x}} & \beta = 3^{-\frac{1}{2}} \\
 \abs{\mathbf{x}}^{-3d/2 + \frac{(\beta^2 + \beta^{-2})d}{4}} \ln^{-\frac{3}{2}}\abs{1/\mathbf{x}} &  \beta \in (3^{-\frac{1}{2}},1] \\
  c' T \abs{\mathbf{x}}^{-2d} \ln^{-\frac{3}{2}}\abs{1/\mathbf{x}} + (1-T) \delta(\mathbf{x}) & \beta > 1 
 \end{dcases} \label{eq:pbpb1}
 \end{equation}

 \end{document}